\documentclass[pra,nofootinbib,floats,aps,twocolumn,tightenlines,superscriptaddress]{revtex4-1}

\usepackage{graphicx}
\usepackage{bbold}
\usepackage{mathtools}
\usepackage{mathrsfs}
\usepackage{hyperref}
\usepackage{commath}
\usepackage{bm}
\usepackage{amsfonts,amsmath,amssymb,amsthm}
\usepackage[usenames,dvipsnames]{xcolor}
\usepackage{subfigure}
\usepackage{dsfont}

\usepackage{upgreek}
\usepackage{qcircuit}
\usepackage[normalem]{ulem}

\DeclareMathOperator{\Tr}{Tr}

\renewcommand*\d[2][]{%
	\mathrm{d}%
	\ifx\relax#1\relax\else
	\rule{-0.02em}{1.5ex}^{#1}\rule{0.08em}{0ex}\!
	\fi
	#2\,
}
\newcommand{\ket}[1]{| {#1} \rangle}
\newcommand{\bra}[1]{\langle {#1} |}
\newcommand{\tr}{\text{Tr}}

\newcommand{\ii}{\mathrm{i}}

\renewcommand{\a}[1]{\hat{a}_{\bm{#1}}}
\newcommand{\ad}[1]{\hat{a}_{\bm{#1}}^\dagger}

\newcommand{\rhoa}{\hat{\rho}_\textsc{a}}
\newcommand{\rhophi}{\hat{\rho}_\phi}
\newcommand{\rhob}{\hat{\rho}_\textsc{b}}
\newcommand{\rhoab}{\hat{\rho}_\textsc{ab}}
\newcommand{\rhoabpt}{\hat{\rho}_\textsc{ab}^{{\text{\textbf{t}}}_\textsc{a}}}

\newcommand{\Hi}{\hat H_\textsc{i}}
\newcommand{\rb}{\hat\rho_{_\beta}}
\newcommand{\om}[1]{\omega_{\bm{#1}}}

\newcommand{\Hphi}{\hat H_\phi}
\newcommand{\D}{\hat D_\alpha}
\renewcommand{\S}{\hat S_\zeta}

\newcommand{\id}{\mathds{1}}

\begin{document}

\title{Harvesting correlations from thermal and squeezed coherent states}
	
\author{Petar Simidzija}	
\email{psimidzija@uwaterloo.ca}
\affiliation{Department of Applied Mathematics, University of Waterloo, Waterloo, Ontario, N2L 3G1, Canada}
\affiliation{Institute for Quantum Computing, University of Waterloo, Waterloo, Ontario, N2L 3G1, Canada}

\author{Eduardo Mart\'in-Mart\'inez}
\email{emartinmartinez@uwaterloo.ca}
\affiliation{Department of Applied Mathematics, University of Waterloo, Waterloo, Ontario, N2L 3G1, Canada}
\affiliation{Institute for Quantum Computing, University of Waterloo, Waterloo, Ontario, N2L 3G1, Canada}
\affiliation{Perimeter Institute for Theoretical Physics, Waterloo, Ontario N2L 2Y5, Canada}

\begin{abstract}

We study the  harvesting of entanglement and mutual information by Unruh-DeWitt particle detectors from thermal and squeezed coherent field states. We prove (for arbitrary spatial dimensions, switching profiles and detector smearings) that while the entanglement harvesting ability of detectors decreases monotonically with the field temperature $T$, harvested mutual information grows linearly with $T$. We also show that entanglement harvesting from a general squeezed coherent state is independent of the coherent amplitude, but depends strongly on the squeezing amplitude. Moreover, we find that highly squeezed states i) allow for detectors to harvest much more entanglement than from the vacuum, and ii) ensure that the entanglement harvested does not decay with their spatial separation. Finally we analyze the spatial inhomogeneity of squeezed states and its influence on harvesting, and investigate how much entanglement one can actually extract from squeezed states when the squeezing is bandlimited.

\end{abstract}
	
\maketitle

\section{Introduction}
\label{sec:intro}

The entanglement structure of a quantum field has been an important area of research over the last few decades. Besides being an interesting focus of study in its own right, the presence of entanglement between local degrees of freedom in general field states (and in particular the vacuum~\cite{Summers1985,Summers1987}) has been used as a means to better understand important fundamental questions, from the black hole information loss problem~\cite{Hawking1975,Hawking1976,Susskind1993,Almheiri2013,Braunstein2013,Hawking2016}, to the dynamics of quantum phase transitions in statistical mechanics~\cite{Vidal2002,Calabrese2004}. Moreover, operational approaches which harness this entanglement to perform useful tasks have also been studied, leading to, for example, the development of protocols for \textit{quantum energy teleportation}~\cite{Hotta2008,Hotta2009,Frey2014}.

Another widely studied protocol making use of the entanglement present in a quantum field is concerned with the extraction of field entanglement onto a pair of initially uncorrelated first-quantized systems (detectors). These so called \textit{entanglement harvesting} protocols were initially studied in the 90s by Valentini~\cite{Valentini1991}, then later by Reznik \textit{et al.}~\cite{Reznik2003,Reznik2005}, and have in the last decade or so experienced a great deal of attention from many different perspectives \cite{Steeg2009,Brown2013a,Martinez2013a,Brown2014,Drago2014,Pozas2015,Salton2015,Pozas2016,Martinez2016a,Nambu,Sachs2017,Guillaume2017,Sachs2018,Trevison2018}.

Many of these recent lines of research into entanglement harvesting are related to the fact that the amount of harvestable entanglement is generally sensitive to the many variable parameters of the setup. For instance, the sensitivity of entanglement harvesting on the position and motion of the detectors has resulted in harvesting-based proposals in metrology --- from rangefiding~\cite{Salton2015} to precise vibration detection~\cite{Brown2014} --- while, on the more fundamental side, it has also been shown that entanglement harvesting is sensitive to the geometry~\cite{Steeg2009} and topology~\cite{Martinez2016a} of the background spacetime. Furthermore, while most of these entanglement harvesting studies have focused on conventional linear Unruh-DeWitt (UDW) particle detectors~\cite{DeWitt1979} coupled to real scalar fields~\cite{Pozas2015}, there have also been several interesting results coming from other variations of the setup. Some examples include: hydrogenoid atomic detectors coupled to the full electromagnetic field~\cite{Pozas2016}, non-linear couplings of UDW detectors to neutral~\cite{Sachs2017} and charged~\cite{Sachs2018} scalar fields, tripartite entanglement in flat spacetime \cite{Drago2014}, and multiple detector harvesting in curved spacetimes \cite{Nambu}. Entanglement harvesting using infinite dimensional harmonic oscillator detectors has been looked at in several works as well. An example which is very relevant to this paper is an article by Brown where the issue of harvesting from thermal states is considered~\cite{Brown2013a}. 

While some of the above mentioned parameters affecting entanglement harvesting are difficult to control in a lab setting (such as the geometry and topology of spacetime), other parameters, such as the energy gap of the detectors or the state of the field, are more easily tunable. A major motivation for studying the sensitivity of entanglement harvesting to these types of parameters is that it may lead to experimental realizations of entanglement harvesting protocols. This would not only be an important achievement from a fundamental perspective, but it could also potentially be a method of obtaining entanglement that could then be used for quantum information purposes~\cite{Martinez2013a}. 

With this ultimate motivation in mind, it has been shown that a non-zero detector energy gap is crucial in protecting an entanglement harvesting UDW pair against fluctuation induced, entanglement harming, local noise~\cite{Pozas2017,Simidzija2018}. Furthermore, for harmonic oscillator detectors, this noise has been found to increase with field temperature, leading to detrimental effects on the amount of entanglement harvested~\cite{Brown2013a} by oscillator pairs. Meanwhile, and perhaps surprisingly, for UDW detectors interacting with coherent states of the field, the presence of leading order local noise does not end up affecting the amount of entanglement that can be harvested from the field~\cite{Simidzija2017b,Simidzija2017c}.

In this paper, we fill in significant gaps in the study of entanglement harvesting sensitivity on thermal and general squeezed coherent field states. While, to our knowledge, this is the first study of squeezed state entanglement harvesting, we would also like to point out that our study of thermal state harvesting differs in several crucial regards to the previous work in~\cite{Brown2013a}. In~\cite{Brown2013a} it was shown that for a pair of pointlike oscillator detectors interacting with a massless field in a one-dimensional cavity, the amount of entanglement extracted decays rapidly with the temperature. In contrast, i) we consider spatially smeared qubit detectors interacting with a field of any mass in a spacetime of any dimensionality, rather than pointlike oscillator detectors interacting with a massless field in (1+1)-dimensions, ii) we look at the continuum free space case rather than being in a cavity, and hence we are not forced to introduce any UV cutoffs to handle numerical sums, and iii) we directly compute the evolved detectors' density matrix from the field's one and two-point functions, rather than using the significantly different formalism of Gaussian quantum mechanics (see, e.g.~\cite{Adesso2007}).

Despite these significant differences between our approach and that in~\cite{Brown2013a}, we will find that, for thermal states, our results are in qualitative agreement with their general conclusions, i.e. that temperature is detrimental to entanglement harvesting. However, since we obtain analytical expressions for entanglement measures, rather than being restricted to numerical calculations, we are able to provide an explicit proof that the amount of entanglement that (qubit) detectors can harvest from the field rapidly decays with its temperature. In particular, we will show that the optimal thermal state for harvesting entanglement from the field is the vacuum. On the other hand, we will see that this is not the case for the harvesting of mutual information, which is a measure of the total (quantum and classical) correlations of the detector pair. In fact we will see that for high field temperatures $T$ (while still in the perturbative regime) the mutual information harvested by the detectors \textit{increases} proportionally with $T$. 

We will then consider the case of squeezed coherent states \cite{Loudon1987}, where, to the authors' knowledge, no previous literature exists. We will first prove that the statement ``entanglement harvesting is independent of the field's coherent amplitude" is true not only for non-squeezed coherent states, as was shown in~\cite{Simidzija2017b}, but also for arbitrarily squeezed coherent states. On the other hand we will show that, unlike the coherent amplitude, the choice of field's squeezing amplitude $\zeta(\bm k)$ does in fact affect the ability of UDW detectors to become entangled, and moreover the Fourier transform of $\zeta(\bm k)$ directly gives the locations in space near which entanglement harvesting is optimal. Perhaps surprisingly, we will also find that for highly and uniformly squeezed field states, the amount of entanglement that the detectors can harvest is independent of their spatial separation, and is often much higher than the amount obtainable from the vacuum. We will also analyze whether this advantage carries over to more experimentally attainable field configurations where states are squeezed across a narrow frequency range of field modes.

This paper is structured as follows: We begin in Sec.~\ref{sec:setup} by reviewing the setup of entanglement harvesting by UDW detectors from arbitrary states of a scalar field. In Sec.~\ref{sec:thermal} we particularize to the case of thermal field states, and study the harvesting of entanglement and mutual information in this setting. Then, in Sec.~\ref{sec:squeezed} we look at entanglement harvesting from squeezed field states, both those with uniform and bandlimited squeezing amplitudes. Finally, Sec.~\ref{sec:conclusions} is left for the conclusions. Units of $\hbar=c=k_\textsc{b}=1$ are used throughout.

\section{Correlation harvesting setup}\label{sec:setup}

Before studying the harvesting of correlations from thermal and squeezed coherent field states, let us review the general correlation harvesting setup that can be found in extensive literature (see, e.g.~\cite{Martinez2015} and references therein) and that is applicable to any field state. We start with a free Klein-Gordon field $\hat\phi$ in $(n+1)$-dimensional Minkowski spacetime, which can be expressed in a basis of plane wave modes as
\begin{equation}
\label{eq:field}
	\hat{\phi}(\bm{x},t)
	=
	\int\frac{\d[n]{\bm{k}}}{\sqrt{2(2\pi)^n \om k}}\left[\ad{k} e^{\ii(\om k t-\bm{k}\cdot\bm{x})}+\text{H.c.}\right],
\end{equation}
where $\om k:=\sqrt{|\bm k|^2+m^2}$, and the creation and annihilation operators, $\ad{k}$ and $\a{k}$, satisfy the canonical commutation relations
\begin{equation}
\label{eq:CCR}
	[\a{k},\a{k'}]= [\ad{k},\ad{k'}]=0, \quad
	[\a{k},\ad{k'}]=\delta^{(n)}(\bm{k}-\bm{k'}).
\end{equation}
We denote by $\ket{0}$ the ground state of the field, by which we mean the state annihilated by all the $\a k$ operators. For now, let us suppose that the field is in an arbitrary (potentially mixed) state $\rhophi$. We will later particularize to the case of thermal and squeezed coherent states.

Next we consider the pair of first-quantized particle detectors that couple to the field with the aim of extracting (i.e. \textit{harvesting}) entanglement. We will model the detectors (labeled $\nu\in\{\text{A},\text{B}\}$) as two-level quantum systems, with ground states $\ket{g_\nu}$, excited states $\ket{g_\nu}$, and proper energy gaps $\Omega_\nu$. We assume that the detectors are at rest at positions $\bm x_\nu$, that they have spatial profiles given by the \textit{smearing functions} $F_\nu(\bm x)$, and that they are initially (i.e. prior to interacting with the field) in the separable state $\rhoa\otimes\rhob$. Then, we describe the interaction of the detectors and the field using the Unruh-DeWitt (UDW) model~\cite{DeWitt1979}, which is a successful model of the light-matter interaction when angular momentum exchange can be neglected~\cite{Pozas2016,Pablo}. In this model the coupling of detectors to field is given by the interaction picture interaction Hamiltonian, $\Hi(t) = \hat H_\textsc{i,a}(t)+\hat H_\textsc{i,b}(t)$, where
\begin{equation}
\label{eq:H_nu}
    \hat{H}_{\textsc{i},\nu}(t)
    :=
    \lambda_\nu \chi_\nu(t) \hat{\mu}_\nu(t)
	\int \d[n]{\bm{x}} F_\nu(\bm{x}-\bm{x}_\nu) \hat{\phi}(\bm{x},t).
\end{equation}
Here, $\lambda_\nu$ is the coupling strength of detector $\nu$ to the field, $\chi_\nu(t)$ is the time-dependent \textit{switching function} which models the duration of the interaction and how the detector $\nu$ is turned on and off, and the $\hat \mu_\nu(t)$ are operators on the two detector Hilbert space given by $
	\hat{\mu}_\textsc{a}(t)
	:=
	\hat{m}_\textsc{a}(t)\otimes\mathds{1}_\textsc{b}$,
and $
    \hat{\mu}_\textsc{b}(t)
	:=		\mathds{1}_\textsc{a}\otimes\hat{m}_\textsc{b}(t)$,
where $\hat{m}_\nu(t)$ is the interaction picture monopole moment of detector $\nu$:
\begin{equation}
    \label{eq:m_nu}
	\hat{m}_\nu(t)=
	\ket{e_\nu}\bra{g_\nu} e^{\ii\Omega_\nu t}+
	\ket{g_\nu}\bra{e_\nu} e^{-\ii\Omega_\nu t}.
\end{equation}

To determine how entangled (if at all) the detectors are following their interactions with the field, we calculate the time-evolved two-detector state $\rhoab$ as
\begin{equation}
\label{eq:rhoab1}
    \rhoab:=\tr_\phi\left[\hat U\left(\rhoa\otimes\rhob\otimes\rhophi\right)\hat U^\dagger\right],
\end{equation}
where the time-evolution unitary $\hat U$ is formally given by
\begin{equation}
\label{eq:U}
	\hat{U}
	=
	\mathcal{T}\exp\left[{-\ii\int_{-\infty}^{\infty}\!\!\!\dif t\, \Hi(t)}\right],
\end{equation}
with $\mathcal T$ denoting the time-ordering operation. By assuming that the detector-field coupling constants $\lambda_\nu$ --- which have units of $(\text{length})^{(n-3)/2}$ in $(n+1)$-dimensional spacetime --- are small compared to other scales with the same units in the setup, we can expand $\hat U$ in powers of $\lambda_\nu$, obtaining
\begin{equation}
\label{eq:Dyson}
	\hat{U}=\id\!
	\underbrace{-\ii\!\int_{-\infty}^\infty\!\!\!\!\!\dif t \Hi(t)} _{\hat{U}^{(1)}}
	\underbrace{-\!\!\int_{-\infty}^{\infty}\!\!\!\!\!\dif t\!
	\int_{-\infty}^t\!\!\!\!\!\dif t' \Hi(t)\hat{H}_\textsc{i,a}(t')}_{\hat{U}^{(2)}}
	+\mathcal{O}(\lambda_\nu^3).
\end{equation}
Then, the final two-detector state $\rhoab$ in Eq.~\eqref{eq:rhoab1} can be perturbatively expressed as
\begin{equation}
\label{eq:rhoab_general}
	\rhoab=
	\hat{\rho}_\textsc{ab}^{(0)}+
	\hat{\rho}_\textsc{ab}^{(1)}+
	\hat{\rho}_\textsc{ab}^{(2)}+
	\mathcal{O}(\lambda_\nu^3),
\end{equation}
where
\begin{align}
    \hat{\rho}_\textsc{ab}^{(0)}&:=
    \rhoa\otimes\rhob\otimes\rhophi,
    \\
	\hat{\rho}_\textsc{ab}^{(1)}&:=
	\tr_\phi\left(\hat{U}^{(1)}\hat{\rho}_0
	+\hat{\rho}_0 \hat{U}^{(1)\dagger}\right),
	\\
	\hat{\rho}_\textsc{ab}^{(2)}&:=
	\tr_\phi\left(\hat{U}^{(2)}\hat{\rho}_0
	+\hat{U}^{(1)} \hat{\rho}_0 
	\hat{U}^{(1)\dagger}
	+\hat{\rho}_0 \hat{U}^{(2)\dagger}\right).
\end{align}
By using the definitions of $\hat U^{(1)}$ and $\hat U^{(2)}$ in Eq.~\eqref{eq:Dyson} and the expression for $\Hi$ given by Eq.~\eqref{eq:H_nu}, it is straightforward to show that $\hat{\rho}_\textsc{ab}^{(1)}$ and $\hat{\rho}_\textsc{ab}^{(2)}$ take the forms
\begin{align}
\label{eq:rhoab^1}
	\hat{\rho}_{\textsc{ab}}^{(1)}&=	\ii\!\!\!\!\sum_{\nu\in\{\text{A,B}\}}\!\!\!\!\lambda_\nu
	\int_{-\infty}^{\infty}\!\!\!\!\!\dif t\chi_\nu(t)
	[\hat{\rho}_{\textsc{ab}}^{(0)},\hat{\mu}_\nu(t)]V(\bm{x}_\nu,t), 
\\
\label{eq:rhoab^2}
	\hat{\rho}_{\textsc{ab}}^{(2)}&=
	\!\!\!\!\sum_{\nu,\eta\in\{\text{A,B}\}}\!\!\!\!
	\lambda_\nu\lambda_\eta
	\Bigg[\int_{-\infty}^{\infty}\!\!\!\!\!\dif t
	\int_{-\infty}^{\infty}\!\!\!\!\!\dif t'
	\chi_\nu(t')\chi_\eta(t) \notag\\
	&\phantom{{}=
	\!\!\!\!\sum_{\nu,\eta\in\{\textsc{a,b}\}}\!\!\!\!
	\lambda_\nu}
	\times\hat{\mu}_\nu(t')
	\hat{\rho}_{\textsc{ab}}^{(0)}\hat{\mu}_\eta(t)
	W(\bm{x}_\eta,t,\bm{x}_\nu,t')\notag\\
	&\phantom{{}=\,\,}
	-\int_{-\infty}^{\infty}\!\!\!\!\!\dif t
	\int_{-\infty}^{t}\!\!\!\!\!\dif t'
	\chi_\nu(t)\chi_\eta(t') \notag\\
	&\phantom{{}=
	\!\!\!\!\sum_{\nu,\eta\in\{\textsc{a,b}\}}\!\!\!\!
	\lambda_\nu}
	\times\hat{\mu}_\nu(t)\hat{\mu}_\eta(t')
	\hat{\rho}_{\textsc{ab}}^{(0)}
	W(\bm{x}_\nu,t,\bm{x}_\eta,t')\notag\\
	&\phantom{{}=\,\,}
	-\int_{-\infty}^{\infty}\!\!\!\!\!\dif t
	\int_{-\infty}^{t}\!\!\!\!\!\dif t'
	\chi_\nu(t)\chi_\eta(t') \notag\\
	&\phantom{{}=
	\!\!\!\!\sum_{\nu,\eta\in\{\textsc{a,b}\}}\!\!\!\!
	\lambda_\nu}
	\times
	\hat{\rho}_{\textsc{ab}}^{(0)}\hat{\mu}_\eta(t')\hat{\mu}_\nu(t)
	W(\bm{x}_\eta,t',\bm{x}_\nu,t)
	\Bigg].
\end{align}
Here, $V(\bm{x}_\nu,t)$ and $W(\bm{x}_\eta,t,\bm{x}_\nu,t')$ are given by
\begin{align}
\label{eq:V}
	V(\bm{x}_\nu,t)
	&\coloneqq
	\int\d[n]{\bm{x}}
	F_\nu(\bm{x}-\bm{x}_\nu)
	v(\bm{x},t),\\
\label{eq:W}
	W(\bm{x}_\eta,t,\bm{x}_\nu,t') &\coloneqq
	\int\!\!\d[n]{\bm{x}}\!\!\!\int\!\!\d[n]{\bm{x'}}
	F_\eta(\bm{x}-\bm{x}_\eta) F_\nu(\bm{x'}-\bm{x}_\nu)
	\notag\\
	&\phantom{{}=}\times w(\bm{x},t,\bm{x'},t'),
\end{align}
while the one- and two-point correlation functions, $v(\bm x,t)$ and $w(\bm x,t,\bm x',t')$, of the field in the state $\rhophi$, are defined as
\begin{align}
	\label{eq:v}
	v(\bm{x},t)
	&:=
	\Tr_\phi
	\left[\hat{\phi}(\bm{x},t)\rhophi\right], 
	\\
	\label{eq:w}
	w(\bm{x},t,\bm{x'},t')
	&:=
	\Tr_\phi
	\left[\hat{\phi}(\bm{x},t)
	\hat{\phi}(\bm{x'},t')
	\rhophi\right].
\end{align}

After computing the evolved two-detector state $\rhoab$ using Eq.~\eqref{eq:rhoab_general}, we can use it to compute the amount of correlations present between the detectors A and B following their interactions with the field. In this paper we will focus on two types of correlations: entanglement and mutual information.

More precisely, we will quantify the entanglement that the detectors A and B harvest from the field by computing the negativity $\mathcal N$, which, for a state $\rhoab$ on the Hilbert space $\mathcal H_\textsc{a}\otimes\mathcal H_\textsc{b}$, is defined as~\cite{Vidal2002}
\begin{equation}
\label{eq:neg}
    \mathcal{N}\left[\rhoab\right]
    \coloneqq
    \sum_i
    \max
    \left(0,-E_{\textsc{ab},i}^{{\text{\textbf{t}}}_\textsc{a}}
    \right),
\end{equation}
where the $E_{\textsc{ab},i}^{{\text{\textbf{t}}}_\textsc{a}}$ are the eigenvalues of the partially transposed matrix $\rhoabpt$. It is well known that the negativity of a two-qubit system is an entanglement monotone that vanishes if and only if the two-qubit state is separable~\cite{Peres1996,Horodecki1996}. Hence the negativity is often used as a measure of entanglement in harvesting scenarios, and it is the measure that we will use.

It is also possible for Alice and Bob to be classically correlated via their interactions with the field. We will quantify the total amount of correlations (quantum and classical)  between them by computing the mutual information, $I$, which is defined as
\begin{align}
\label{eq:mut_info}
    I[\rhoab]:=
    S[\rhoa]+S[\rhob]-S[\rhoab],
\end{align}
where $S[\hat\rho]:=-\tr(\hat\rho\log\hat\rho)$ is the von Neumann entropy of the state $\hat\rho$, while $\rhoa:=\tr_\textsc{b}(\rhoab)$ and $\rhob:=\tr_\textsc{a}(\rhoab)$ are the reduced states of detectors A and B following the detector-field interactions. In particular, if entanglement is zero and the mutual information is not, the correlations have to be either classical correlations or discord \cite{Ollivier2001,Henderson2001}.

\section{Thermal field state}
\label{sec:thermal}

Let us suppose now that the two Unruh-DeWitt detectors are initially in their ground states, $\hat\rho_\nu=\ket{g_\nu}\bra{g_\nu}$, and that the field is in a thermal state $\rb$ of inverse temperature $\beta$. It will be sufficient for our purposes to formally define $\rb$ as a Gibbs state in the usual way. Namely we write
\begin{equation}
\label{eq:rb}
    \rb:=\frac{\exp(-\beta\hat H_\phi)}{Z},
\end{equation}
where $Z:=\tr[\exp(-\beta\hat H_\phi)]$ is the partition function of the free field. Here $\hat H_\phi$ is the Shr\"odinger picture free field Hamiltonian, which, after subtracting off an infinite zero-point energy (which does not affect any observable dynamics), takes the form
\begin{equation}
    \hat H_\phi = \int \d[n]{\bm k} \om{k}
    \ad{k}\a{k}.
\end{equation}

We would like to emphasize that, strictly speaking, the Gibbs definition of $\rb$ in Eq.~\eqref{eq:rb} is not well defined when $\hat H_\phi$ is the Hamiltonian of a field in free space, since then $\hat H_\phi$ is an operator acting on a Hilbert space of uncountably many dimensions, and certain technical issues arise in with performing its exponentiation and trace. We could proceed rigorously by instead considering our field to be in a large box of length $L$, such that its Hilbert space is of countable dimension, and then in the end taking the limit $L\rightarrow\infty$. Alternatively we could formalize our treatment by making use of the Kubo-Martin-Schwinger (KMS) definition of a thermal state, which is rigorously defined even for continuous variable systems~\cite{Kubo1957,Martin1959}. In this case the definition of $\rb$ would correspond to a KMS state of KMS parameter $\beta$ with respect to the time $t$ proper to both detectors. However we will shortly see that, for our limited purposes, these more rigorous definitions of $\rb$ are unnecessary in the sense that formal calculations using the Gibbs definition in Eq.~\eqref{eq:rb} yield the same results. This can be checked by comparing the results we will obtain with, e.g.,~\cite{Strocchi2008}.

To see this concretely, from the definition~\eqref{eq:rb} of $\rb$ and the canonical commutation relations (CCRs) in Eq.~\eqref{eq:CCR}, we can straightforwardly calculate the one- and two-point correlation functions defined in~\eqref{eq:v} and \eqref{eq:w}. Because the field is composed of a linear superposition of $\a k$ and $\ad k$ operators, we first compute the following useful expression:
\begin{align}\label{eq:expectation_a}
    \tr_\phi\left(\rb\a k\right)
    &=
    \frac{1}{Z}\tr_\phi\left(e^{-\beta\Hphi}\a k\right)\\
    &=
    \frac{1}{Z}\tr_\phi\left(e^{-\beta\Hphi}\a k e^{\beta\Hphi}e^{-\beta\Hphi}\right)\notag\\
    &=
    \frac{e^{\beta\om k}}{Z}\tr_\phi\left(\a k e^{-\beta\Hphi}\right)\notag\\
    &=
    e^{\beta\om k}\tr_\phi\left(\rb\a k\right)\notag,
\end{align}
where in the third line we made use of the identity $e^{-\beta\Hphi}\a k e^{\beta\Hphi}=e^{\beta\om k}\a k$, which can be easily proved using the Zassenhaus formula and the CCRs. Then, comparing the first and last lines of Eq.~\eqref{eq:expectation_a}, we conclude that $\tr_\phi\left(\rb\a k\right)=0$. Hence $\tr_\phi(\rb\ad k)=0$, and therefore  the one-point function $v(\bm x,t)=0$. Then, from Eqs.~\eqref{eq:rhoab^1} and \eqref{eq:V}, we conclude that the first order contribution $\hat\rho_\textsc{ab}^{(1)}$ to $\rhoab$ is identically zero for a thermal field state. 

To calculate the two-point function $w(\bm x,t,\bm x',t')$ we first compute:
\begin{align}\label{eq:expectation_a_ad}
    \tr_\phi\left(\rb\a k\ad{k'}\right)
    &=
    \frac{1}{Z}\tr_\phi\left(e^{-\beta\Hphi}\a k\ad{k'}\right)\\
    &=
    \frac{1}{Z}\tr_\phi\left(e^{-\beta\Hphi}\a k e^{\beta\Hphi}e^{-\beta\Hphi\ad{k'}}\right)\notag\\
    &=
    \frac{e^{\beta\om k}}{Z}\tr_\phi\left(\a k e^{-\beta\Hphi\ad{k'}}\right)\notag\\
    &=
    e^{\beta\om k}\tr_\phi\left(\rb\ad{k'}\a k\right)\notag\\
    &=e^{\beta\om k}\left[\tr_\phi\left(\rb\a k\ad{k'}\right)+\delta(\bm k-\bm k')\right]\notag,
\end{align}
where in the last step we again made use of the CCRs. Comparing the first and last lines of this expression gives the result
\begin{equation}
\label{eq:exp_a_ad}
    \tr(\rb\a{k}\ad{k'})=
    \frac{e^{\beta\om{k}}}{e^{\beta\om{k}}-1}\delta^3(\bm k-\bm k').
\end{equation}
Similarly we obtain the identities
\begin{align}
    \tr(\rb\ad{k}\a{k'})
    &=
    \frac{1}{e^{\beta\om{k}}-1}\delta^3(\bm k-\bm k'),\\
    \tr(\rb\a{k}\a{k'})
    &=0,\\
    \tr(\rb\ad{k}\ad{k'})
    &=0.
\label{eq:exp_ad_ad}
\end{align}

Notice that, as alluded to above, the calculations in Eqs.~\eqref{eq:expectation_a} and \eqref{eq:expectation_a_ad} would turn out the same if we rigorously considered the field in a box and then took the $L\rightarrow\infty$ limit in the end. In particular the only difference would be that the CCRs contain a Kronecker delta, which in the limit of free space becomes a Dirac delta, thus recovering our results in a more rigorous fashion. Furthermore, our final expressions in Eqs.~\eqref{eq:exp_a_ad}-\eqref{eq:exp_ad_ad} are equal to those obtained using the KMS definition of $\rb$ (see equation 14.3 in~\cite{Strocchi2008}). Hence our formal use of the Gibbs definition of $\rb$ in Eq.~\eqref{eq:rb} is justified.

We can now use the identities in Eqs.~\eqref{eq:exp_a_ad}-\eqref{eq:exp_ad_ad} to write the two-point function of the field, defined by $w(\bm x,t,\bm x',t'):=\tr[\rb\hat\phi(\bm x,t)\hat\phi(\bm x',t')]$, as
\begin{align}\label{eq:thermal_wightman}
    w(\bm x,t,\bm x',t')
    =
    w^\text{vac}(\bm x,t,\bm x',t')
    +w^\text{th}_\beta(\bm x,t,\bm x',t').
\end{align}
Here $w^\text{vac}(\bm x,t,\bm x',t')$ and $w^\text{th}_\beta(\bm x,t,\bm x',t')$ are the vacuum ($\beta$-independent) two-point function  and the thermal ($\beta$-dependent) contribution, respectively, and are explicitly given by
\begin{align}
    \label{eq:wvac}
    w^\text{vac}(\bm x,t,\bm x',t')&=\!
    \int \!\frac{\d[n]{\bm k}}{2(2\pi)^n\om{k}}
        e^{-\ii\om{k}(t-t')}
        e^{\ii\bm k\cdot(\bm x-\bm x')},
    \\
    \label{eq:wbeta}
    w^\text{th}_\beta(\bm x,t,\bm x',t')&=
    \int \frac{\d[n]{\bm k}\left[
    e^{\ii\om{k}(t-t')}
    e^{-\ii\bm k\cdot(\bm x-\bm x')}
    +\text{c.c}
    \right]}{2(2\pi)^n\om{k}\left(e^{\beta\om{k}}-1\right)}
    .
\end{align}

Before we proceed to use the two-point function to calculate the time-evolved two-detector density matrix $\rhoab$, it should be noted that in the literature one often finds a very different looking expression for the two-point function of a thermal field state. For instance, in~\cite{Weldon2000}, the thermal two-point function for a massless field in $(3+1)$-dimensions is shown to be
\begin{align}\label{eq:thermal_wightman_alt}
    w(\bm x, t, 0, 0)=&
    \frac{1}{8\pi r\beta}
    \!\left[\coth\!\left(\frac{\pi(r+t)}{\beta}\right)\!+\coth\!\left(\frac{\pi(r-t)}{\beta}\right)\!
    \right]\notag\\
    &+
    \frac{\ii}{8\pi r}
    \left[\delta^{(3)}(r+t)-\delta^{(3)}(r-t)]
    \right],
\end{align}
where $r:=|\bm x|$. The advantage of this expression over the one in Eq.~\eqref{eq:thermal_wightman} is that there are no integrals over momentum space that have to be evaluated. The disadvantage is that it is restrictive to the massless $(3+1)$-dimensional case. Furthermore the method used in \cite{Weldon2000} to obtain Eq.~\eqref{eq:thermal_wightman_alt} is much less direct than the method we employed in obtaining Eq.~\eqref{eq:thermal_wightman}. In any case, as a consistency check in Appendix~\ref{app:thermal_wightman} we show that the expression in Eq.~\eqref{eq:thermal_wightman_alt} is indeed a specific case of Eq.~\eqref{eq:thermal_wightman} when $m=0$, $n=3$, and $\bm x' = t' = 0$.

We now come back to our main objective: use the two-point function $w(\bm x,t,\bm x',t')$ in Eq.~\eqref{eq:thermal_wightman} to compute the density matrix $\rhoab$ in \eqref{eq:rhoab1}. Substituting \eqref{eq:thermal_wightman} into \eqref{eq:rhoab^2} we obtain
\begin{equation}
	\label{eq:rhoab_thermal}
    \rhoab=
    \begin{pmatrix}
	1-\mathcal{L}_\textsc{aa}(\beta)-\mathcal{L}_\textsc{bb}(\beta)& 0 & 0 & \mathcal{M}^*(\beta) \\
	0 & \mathcal{L}_\textsc{bb}(\beta) & \mathcal{L}_\textsc{ab}^*(\beta) & 0 \\
	0 & \mathcal{L}_\textsc{ab}(\beta) & \mathcal{L}_\textsc{aa}(\beta) & 0 \\
	\mathcal{M}(\beta) & 0 & 0 & 0
	\end{pmatrix},
\end{equation}
to second order in the coupling strength $\lambda$, and where we work in the basis $\{    \ket{g_\textsc{a}}\ket{g_\textsc{b}},
\ket{g_\textsc{a}}\ket{e_\textsc{b}},
\ket{e_\textsc{a}}\ket{g_\textsc{b}},
\ket{e_\textsc{a}}\ket{e_\textsc{b}}\}$. The terms $\mathcal{L}_{\nu\eta}(\beta)$ and $\mathcal M(\beta)$ are defined to be
\begin{align}
    \label{eq:L_thermal}
    \mathcal{L}_{\nu\eta}(\beta)&=
    \mathcal{L}_{\nu\eta}^\text{vac}
    +2\pi\lambda_\nu\lambda_\eta
    \int\!\!
    \frac{\d[n]{\bm k}\bar F_\nu^*(\bm k) \bar F_\eta(\bm k)  e^{\ii\bm k\cdot(\bm x_\eta-\bm x_\nu)}}{2\om{k}\left(e^{\beta\om{k}}-1\right)}\notag\\
    &\hspace{1.5cm}\times\Big[
    \bar \chi_\nu^*(\om k-\Omega_\nu)
    \bar \chi_\eta(\om k-\Omega_\eta)
    \notag\\
    &\hspace{2cm}+
    \bar \chi_\nu(\om k+\Omega_\nu)
    \bar \chi_\eta^*(\om k+\Omega_\eta)
    \Big],
    \\
    \label{eq:M_thermal}
    \mathcal M(\beta)&=
    \mathcal M^\text{vac}
    -2\pi\lambda_\textsc{a}\lambda_\textsc{b}
    \int\!\!
    \frac{\d[n]{\bm k}\bar F_\textsc{a}(\bm k)\bar F_\textsc{b}^*(\bm k) e^{\ii\bm k\cdot(\bm x_\textsc{a}-\bm x_\textsc{b})}}{2\om{k}\left(e^{\beta\om{k}}-1\right)}\notag\\
    &\hspace{1.5cm}\times\Big[
    \bar \chi_\textsc{a}^*(\om k-\Omega_\textsc{a})
    \bar \chi_\textsc{b}(\om k+\Omega_\textsc{b})
   \notag\\
    &\hspace{2cm}+\bar \chi_\textsc{a}(\om k+\Omega_\textsc{a})
    \bar \chi_\textsc{b}^*(\om k-\Omega_\textsc{b})
    \Big].
\end{align}
Here we define the Fourier transform $\bar g:\mathbb R^m\rightarrow\mathbb C$ of a function $g:\mathbb R^m\rightarrow\mathbb R$ as
\begin{equation}
	\label{eq:FT}
    \bar{g}(\bm{k})
	:=
	\frac{1}{\sqrt{(2\pi)^m}}\int\d[m]{\bm{x}}
	g(\bm{x})e^{\ii\bm{k}\cdot\bm{x}},
\end{equation}
and as always we use the superscript ``vac" to denote quantities that do not depend on the inverse temperature $\beta$, i.e. those terms which arise from the ``vacuum" part $w^\text{vac}$ of the two-point function. The vacuum terms $\mathcal{L}_{\nu\eta}^\text{vac}$ and $\mathcal M^\text{vac}$ are explicitly given by
\begin{align}
    \label{eq:Lvac}
    \mathcal L_{\nu\eta}^\text{vac}
    &=
    2\pi\lambda_\nu\lambda_\eta
    \int\frac{\d[n]{\bm k}}{2\om k}
    \bar F_\nu^*(\bm k)\bar F_\eta(\bm k)
    e^{-\ii\bm k\cdot(\bm x_\nu-\bm x_\eta)}
    \\
    &\hspace{2cm}\times
    \bar \chi_\nu(\om k+\Omega_\nu)
    \bar \chi_\eta^*(\om k+\Omega_\eta),
    \notag\\
    \label{eq:Mvac}
    \mathcal M^\text{vac}
    &=
    -\lambda_\textsc{a}\lambda_\textsc{b}
    \int\frac{\d[n]{\bm k}}{2\om k}
    \int_{-\infty}^\infty\dif t
    \int_{-\infty}^t\dif t'
    e^{-\ii\om k(t-t')}
    \\
    &\hspace{1cm}\times
    \Big[
    \bar F_\textsc{a}(\bm k)
    \bar F_\textsc{b}^*(\bm k)
    e^{\ii\bm k\cdot(\bm x_\textsc{a}-\bm x_\textsc{b})}
    \bar \chi_\textsc{a}(t)
    \bar \chi_\textsc{b}(t')
    \notag\\
    &\hspace{1.5cm}\times
    e^{\ii(\Omega_\textsc{a}t+\Omega_\textsc{b}t')}
    +
    (\text{A}\leftrightarrow\text{B})
    \Big].
    \notag
\end{align}

\subsection{Harvesting entanglement}
\label{eq:thermal:harvesting_entanglement}

Having computed the time-evolved density matrix $\rhoab$ of the Unruh-DeWitt detector pair, we can now compute the negativity of this state and thus quantify the amount of entanglement the detectors harvest from the thermal field state. Using the expression \eqref{eq:rhoab_thermal} for $\rhoab$, we find that in the same computational basis, to $\mathcal O(\lambda^2)$ the partially transposed matrix $\rhoabpt$ takes the form
\begin{equation}
	\label{eq:rhoab_thermal_pt}
    \rhoabpt=
    \begin{pmatrix}
	1-\mathcal{L}_\textsc{aa}(\beta)-\mathcal{L}_\textsc{bb}(\beta)& 0 & 0 & \mathcal{L}_\textsc{ab}^*(\beta) \\
	0 & \mathcal{L}_\textsc{bb}(\beta) &  \mathcal{M}^*(\beta) & 0 \\
	0 & \mathcal{M}(\beta) & \mathcal{L}_\textsc{aa}(\beta) & 0 \\
	\mathcal{L}_\textsc{ab}(\beta) & 0 & 0 & 0
	\end{pmatrix}.
\end{equation}
As discussed in~\cite{Pozas2015}, at $\mathcal O(\lambda^2)$ a matrix of this form has only one potentially negative eigenvalue:
\begin{align}
\label{eq:neg2}
    E_{\textsc{ab},1}^{{\text{\textbf{t}}}_\textsc{a}}
    =
    \frac{1}{2}\Big(&
	\mathcal{L}_\textsc{aa}(\beta)+\mathcal{L}_\textsc{bb}(\beta)\\
	&-\sqrt{(\mathcal{L}_\textsc{aa}(\beta)-\mathcal{L}_\textsc{bb}(\beta))^2+
	4|\mathcal{M}(\beta)|^2}\Big).\notag
\end{align}
Hence we find that the negativity $\mathcal N$, defined in Eq.~\eqref{eq:neg}, can be written as
\begin{equation}
    \mathcal N[\rhoab]=\max\left(0,-E_{\textsc{ab},1}^{{\text{\textbf{t}}}_\textsc{a}}\right).
\end{equation}

Now suppose that the detectors A and B are identical. That is, they have the same shapes $F(\bm x)=F_\nu(\bm x)$, the same proper energy gaps $\Omega=\Omega_\nu$, the same coupling constants $\lambda=\lambda_\nu$, and the same switching profiles $\chi(t-t_\nu)=\chi_\nu(t)$. Note that we are still allowing for the detectors to couple to the field at potentially different spacetime locations $(t_\textsc{a},\bm x_\textsc{a})$ and $(t_\textsc{b},\bm x_\textsc{b})$. However, since the local terms $\mathcal{L}_{\nu\nu}$ are translationally invariant, we find that $\mathcal L_\textsc{aa}(\beta)=\mathcal L_\textsc{bb}(\beta)$, and the negativity can be written more simply as
\begin{equation}
\label{eq:neg_thermal}
    \mathcal N=\max\left[0,|\mathcal M(\beta)|-\mathcal L_{\nu\nu}(\beta)\right].
\end{equation}
As acknowledged in \cite{Pozas2015}, this form for the negativity makes evident the competition between the non-local term $|\mathcal M(\beta)|$, which increases the negativity, and the local term $\mathcal L_{\nu\nu}(\beta)$, which decreases it. We note however, that although this interpretation of Eq.~\eqref{eq:neg_thermal} is pleasantly consistent with the intuition that entanglement is a non-local phenomenon, it should not be taken too literally. For instance, in~\cite{Simidzija2017b,Simidzija2017c} it was shown that a detector pair interacting with a coherent field state extracts the exact same amount of entanglement as it would from a vacuum state, despite the fact that inherently local terms of $\mathcal O(\lambda)$ appear in $\rhoab$ for the former but not the latter case.

Having obtained an expression in \eqref{eq:neg_thermal} for the negativity $\mathcal N$ of two identical Unruh-DeWitt detectors following their interactions with a thermal field state, we would now like to determine the temperature dependence of $\mathcal N$. In other words, we want to answer the question, ``what is the optimal field temperature for Unruh-DeWitt detectors to harvest entanglement?"

To answer this question, let us first particularize the terms $\mathcal L_{\nu\eta}(\beta)$ and $\mathcal M(\beta)$ in Eqs.~\eqref{eq:L_thermal} and \eqref{eq:M_thermal} for identical detectors. We obtain
\begin{align}
    \label{eq:L_thermal_identical}
    \mathcal{L}_{\nu\eta}(\beta)&=
    \mathcal{L}_{\nu\eta}^\text{vac}
    +\pi\lambda^2
    \int\!\!
    \frac{\d[n]{\bm k}|\bar F(\bm k)|^2}{\om{k}\left(e^{\beta\om{k}}-1\right)}
    e^{\ii\bm k\cdot(\bm x_\eta-\bm x_\nu)}
    \notag
    \\
    &\hspace{0.3cm}\times
    \Big(
    |\bar \chi(\om k-\Omega)|^2
    e^{\ii(\om k-\Omega)t_\eta}
    e^{-\ii(\om k-\Omega)t_\nu}
    \notag\\
    &\hspace{0.7cm}
    + |\bar \chi(\om k+\Omega)|^2
    e^{-\ii(\om k+\Omega)t_\eta}
    e^{\ii(\om k+\Omega)t_\nu}
    \Big),
    \\
    \label{eq:M_thermal_identical}
    \mathcal M(\beta)&=
    \mathcal M^\text{vac}
    -2\pi\lambda^2\!\!
    \int\!\!
    \frac{\d[n]{\bm k}|\bar F(\bm k)|^2
    e^{\ii\Omega(t_\textsc{a}+t_\textsc{b})}
    e^{\ii\bm k\cdot(\bm x_\textsc{a}-\bm x_\textsc{b})}}{\om{k}\left(e^{\beta\om{k}}-1\right)}\notag
    \\
    &\hspace{0.3cm}
    \times
    \bar \chi^*(\om k-\Omega)
    \bar \chi(\om k+\Omega)
    \cos[\om k(t_\textsc{a}-t_\textsc{b})].
\end{align}
Now, let us consider two temperatures, $\beta_1^{-1}<\beta_2^{-1}$. Then, defining
\begin{equation}
\label{eq:h}
    h(\bm k):=
    \frac{1}{e^{\beta_2\om k}-1}
    -
    \frac{1}{e^{\beta_1\om k}-1},
\end{equation}
which is strictly greater than zero, we can rewrite $\mathcal L_{\nu\nu}(\beta)$ and $\mathcal M(\beta)$ to read
\begin{align}
    \label{eq:L_thermal_identical_2}
    \mathcal{L}_{\nu\nu}(\beta_2)&=
    \mathcal{L}_{\nu\nu}(\beta_1)
    +\pi\lambda^2
    \int\!\!
    \frac{\d[n]{\bm k}h(\bm k)|\bar F(\bm k)|^2}{\om{k}}
    \notag
    \\
    &\hspace{1cm}\times
    \Big(
    |\bar \chi(\om k-\Omega)|^2
    + |\bar \chi(\om k+\Omega)|^2
    \Big),
    \\
    \label{eq:M_thermal_identical_2}
    \mathcal M(\beta_2)&=
    \mathcal M(\beta_1)
    -2\pi\lambda^2\!\!
    \int\!\!
    \frac{\d[n]{\bm k}h(\bm k)|\bar F(\bm k)|^2
    }{\om{k}}\notag
    \\
    &\hspace{1cm}
    \times
    e^{\ii\Omega(t_\textsc{a}+t_\textsc{b})}
    e^{\ii\bm k\cdot(\bm x_\textsc{a}-\bm x_\textsc{b})}
    \cos[\om k(t_\textsc{a}-t_\textsc{b})]\notag
    \\
    &\hspace{1cm}
    \times
    \bar \chi^*(\om k-\Omega)
    \bar \chi(\om k+\Omega).
\end{align}
Taking the magnitude of the latter expression we obtain
\begin{align}
    \label{eq:M_thermal_norm}
    |\mathcal M(\beta_2)|
    &\le
    |\mathcal M(\beta_1)|
    +2\pi\lambda^2
    \Bigg|
    \int\!\!
    \frac{\d[n]{\bm k}h(\bm k)|\bar F(\bm k)|^2
    }{\om{k}}\notag
    \\
    &\hspace{1cm}
    \times
    e^{\ii\Omega(t_\textsc{a}+t_\textsc{b})}
    e^{\ii\bm k\cdot(\bm x_\textsc{a}-\bm x_\textsc{b})}
    \cos[\om k(t_\textsc{a}-t_\textsc{b})]\notag
    \\
    &\hspace{1cm}
    \times
    \bar \chi^*(\om k-\Omega)
    \bar \chi(\om k+\Omega)
    \Bigg|
    \notag
    \\
    &\le
    |\mathcal M(\beta_1)|
    +2\pi\lambda^2
    \int\!\!
    \frac{\d[n]{\bm k}h(\bm k)|\bar F(\bm k)|^2
    }{\om{k}}\notag
    \\
    &\hspace{1cm}
    \times
    |\bar \chi^*(\om k-\Omega)|
    |\bar \chi(\om k+\Omega)|
\end{align}
Finally, combining Eqs.~\eqref{eq:L_thermal_identical_2} and \eqref{eq:M_thermal_norm} we find
\begin{align}
\label{eq:thermal_ineq}
    &\,\,
    |\mathcal M(\beta_2)|-\mathcal L_{\nu\nu}(\beta_2)\notag
    \\
    \le&\,\,
    |\mathcal M(\beta_1)|-\mathcal L_{\nu\nu}(\beta_1)
    -\pi\lambda^2\!\!
    \int\!\!
    \frac{\d[n]{\bm k}h(\bm k)D(\bm k)}{\om{k}}
    \notag\\
    \le&\,\,
    |\mathcal M(\beta_1)|-\mathcal L_{\nu\nu}(\beta_1),
\end{align}
where $D(\bm k):=|\bar F(\bm k)|^2(
    |\bar \chi(\om k-\Omega)|\!-\!
    |\bar \chi(\om k+\Omega)|
    )^2$
is a non-negative function characterized by the switching, smearing, and energy gap of the detectors. Hence, using the definition \eqref{eq:neg2} of the negativity, Eq.~\eqref{eq:thermal_ineq} proves our first result: the amount of entanglement that two identical UDW detectors can harvest from a thermal field state decreases with the temperature $\beta^{-1}$. This is true regardless of the dimensionality of spacetime, the mass of the field, and the properties (spatial smearing, temporal switching, energy gap) of the detectors. 

In fact, we can obtain a somewhat stronger statement about the negativity of a pair of detectors interacting with a thermal field state. First, notice from Eq.~\eqref{eq:h} that for given values of $\beta_1$ and $\bm k$, the value of the function $h(\bm k)$ can be increased arbitrarily by choosing a small enough value of $\beta_2$. Therefore, from Eq.~\eqref{eq:thermal_ineq}, as long as $D(\bm k)$ is not identically equal to zero, we find that the value of $|\mathcal M(\beta_2)|-\mathcal L_{\nu\nu}(\beta_2)$ can be made negative by taking a large enough temperature $\beta_2^{-1}$. Hence, not only does the amount of entanglement harvested by a UDW detector pair decreases monotonically with the temperature, but also by increasing the temperature of the field to a high enough value we can always (as long as $D(\bm k)$ is not identically zero) ensure that the thermal noise prevents the detectors from becoming entangled at all. This is true regardless the mass of the field, spacetime dimensionality and the detector properties.

Knowing that the negativity $\mathcal N$ of a detector pair decreases with the temperature of the field, we can ask what is the rate of this decrease. We can straightforwardly obtain a bound on $\dif \mathcal N/\dif \beta$ from Eq.~\eqref{eq:thermal_ineq}. First, writing $E_{\textsc{ab},1}^{{\text{\textbf{t}}}_\textsc{a}}(\beta)=\mathcal L_{\nu\nu}(\beta)-|\mathcal M(\beta)|$ for identical detectors, the second line of Eq.~\eqref{eq:thermal_ineq} can be expressed as
\begin{align}
    E_{\textsc{ab},1}^{{\text{\textbf{t}}}_\textsc{a}}(\beta_1)
    -
    E_{\textsc{ab},1}^{{\text{\textbf{t}}}_\textsc{a}}(\beta_2)
    \le
    -\pi\lambda^2\!\!
    \int\!\!
    \frac{\d[n]{\bm k}h(\bm k)D(\bm k)}{\om{k}}.
\end{align}
Dividing both sides of this expression by $\beta_1-\beta_2$, taking the limit $\beta_1\rightarrow\beta_2$, and using the fact that
\begin{equation}
    \lim_{\beta_1\rightarrow\beta_2}
    \frac{h(\bm k)}{\beta_1-\beta_2}
    =
    -\frac{\dif}{\dif\beta_1}
    \left(
    \frac{1}{e^{\beta_1\om k}-1}
    \right)
    =
    \frac{\om k e^{\beta_1\om k}}{\left(e^{\beta_1\om k}-1\right)^2},
\end{equation}
we find the rate of change of the eigenvalue $E_{\textsc{ab},1}^{{\text{\textbf{t}}}_\textsc{a}}(\beta)$ with respect to the inverse temperature $\beta$ to be bounded from below according to
\begin{align}
    \frac{\dif}{\dif\beta}
    E_{\textsc{ab},1}^{{\text{\textbf{t}}}_\textsc{a}}(\beta)
    &\le
    -\pi\lambda^2\!\!
    \int
    \d[n]{\bm k}D(\bm k)
    \frac{e^{\beta\om k}}{\left(e^{\beta\om k}-1\right)^2}
    .
\end{align}
Therefore in regions where the negativity $\mathcal N(\beta)$ is non-zero, we have that
\begin{equation}
    \frac{\dif\mathcal N}{\dif\beta}
    \ge
    \pi\lambda^2\!\!
    \int
    \d[n]{\bm k}D(\bm k)
    \frac{e^{\beta\om k}}{\left(e^{\beta\om k}-1\right)^2}.
\end{equation}
This puts a lower bound on how fast $\mathcal N$ must grow with the inverse temperature $\beta$, in regions where $\mathcal N$ is non-zero. Of course if $\mathcal N$ is zero, then increasing $\beta$ will only result in $\mathcal N$ remaining zero. 

Having proven the general result that temperature is always detrimental to entanglement harvesting (at least for identical detectors), let us now consider some particular parameters for the detectors A and B, so that we may explicitly see the manifestation of this phenomenon. To that end, let us suppose that the two detectors are located in $(3+1)$ dimensional spacetime, that they have Gaussian spatial profiles of width $\sigma$,
\begin{align}
\label{eq:smearing_gaussian}
    F(\bm x)=\frac{1}{(\sqrt{\pi}\sigma)^3}
    e^{-\frac{|\bm x|^2}{\sigma^2}},
\end{align}
and that their temporal switching functions are also Gaussians (of width $\uptau$),
\begin{align}
\label{eq:switching_gaussian}
    \chi(t)=e^{-\frac{t^2}{\uptau^2}}.
\end{align}
Then it is straightforward to show that the terms $\mathcal M^\text{th}$, $\mathcal L_\textsc{ab}^\text{th}$ and $L_{\nu\nu}^\text{th}$, which make up the thermal contributions to the density matrix $\rhoab$, evaluate to
\begin{align}
    \mathcal M^\text{th}&=
    -\frac{\tilde\lambda^2e^{-\frac{1}{2}\tilde\Omega^2}e^{\ii\tilde\Omega\tilde\Delta^+}}{4\pi\tilde d}
    \int_0^\infty\dif \tilde k
    \frac{e^{-\frac{1}{2}\tilde k^2(1+\tilde\sigma^2)}}{e^{\tilde\beta\tilde k}-1}
    \\
    \notag
    &\hspace{3cm}\times
    \sin(\tilde d\tilde k)\cos(\tilde\Delta^-\tilde k),
    \\
    \mathcal L_\textsc{ab}^\text{th}&=
    \frac{\tilde\lambda^2e^{-\frac{1}{2}\tilde\Omega^2}e^{-\ii\tilde\Omega\tilde\Delta^-}}{2\pi\tilde d}
    \int_0^\infty\dif \tilde k
    \frac{e^{-\frac{1}{2}\tilde k^2(1+\tilde\sigma^2)}}{e^{\tilde\beta\tilde k}-1}
    \\
    \notag
    &\hspace{3cm}\times
    \sin(\tilde d\tilde k)\cosh[(\tilde\Omega+\ii\tilde\Delta^-)\tilde k],
    \\
    \mathcal L_{\nu\nu}^\text{th}&=
    \frac{\tilde\lambda^2e^{-\frac{1}{2}\tilde\Omega^2}}{2\pi}
    \int_0^\infty\dif \tilde k
    \frac{\tilde k e^{-\frac{1}{2}\tilde k^2(1+\tilde\sigma^2)}}{e^{\tilde\beta\tilde k}-1}
    \cosh(\tilde\Omega\tilde k).
\end{align}
Here, every quantity with a tilde is a dimensionless expression of the scales of the problem in units of $\uptau$ (e.g. $\tilde\Omega:=\Omega \uptau$, $\tilde\beta:=\beta/\uptau$), and we have defined $\tilde d:=|\bm{ x}_\textsc{a}-\bm{ x}_\textsc{b}|/\uptau$ and $\tilde \Delta^\pm:= (t_\textsc{b}\pm t_\textsc{a})/\uptau$. Meanwhile the terms $\mathcal M^\text{vac}$ and $\mathcal L_{\nu\eta}^\text{vac}$, which give the vacuum ($\beta$ independent) contributions to $\rhoab$, can be found in equations 29-31 in~\cite{Pozas2015}.

\begin{figure}
    \centering
    \includegraphics[width=0.9\linewidth]{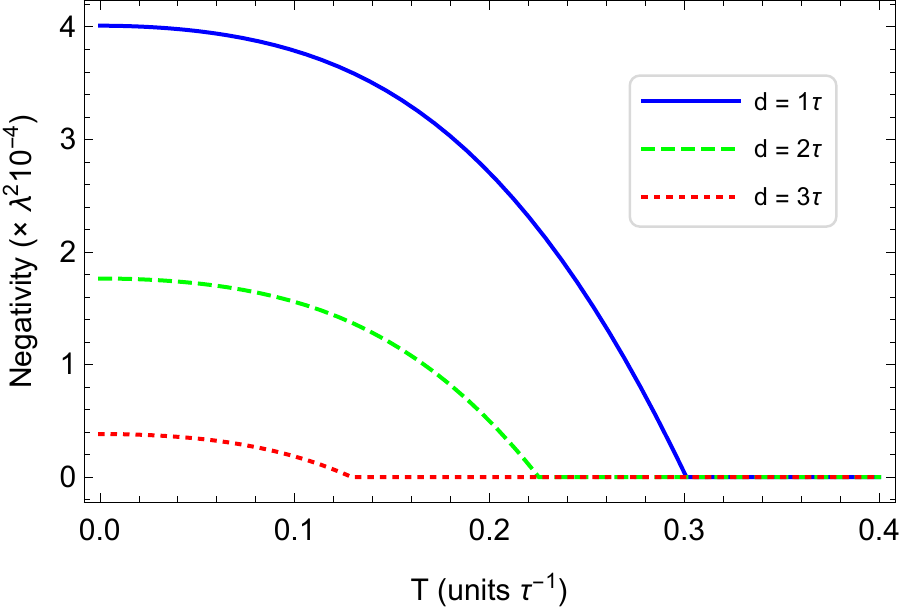}
    \caption{Negativity of identical detectors as a function of field temperature, for different spatial separations $d$ of their centers of mass. The detectors are coupled to the field at the same time according to a Gaussian switching function of width $\uptau$, their spatial profiles are Gaussians of width $\sigma=\uptau$, and their energy gap is $\Omega=3/\uptau$.}
    \label{fig:NegVsT_thermal}
\end{figure}

Assuming these detector spatial profiles and switching functions, in Fig.~\ref{fig:NegVsT_thermal} we show the dependence of the negativity of the detector pair on the temperature $T=\beta^{-1}$ of the field. We see that, in accordance with our general discussion above, the negativity is a monotonically decreasing function of $T$, and that it is identically zero after a certain finite temperature. These findings are qualitatively the same as what was found in~\cite{Brown2013a}, namely that harmonic oscillator detectors in a (1+1)D cavity harvest less entanglement as the field temperature increases. This is, of course, all in agreement with our intuition that ``thermal noise" is detrimental to the detectors obtaining non-local correlations. We will soon see however, that this seemingly reasonable intuition does not apply when we quantify the correlations using the mutual information rather than the negativity. In particular we will show that the mutual information between the detector pair can increase with the field temperature.

\begin{figure}
    \centering
    \includegraphics[width=0.9\linewidth]{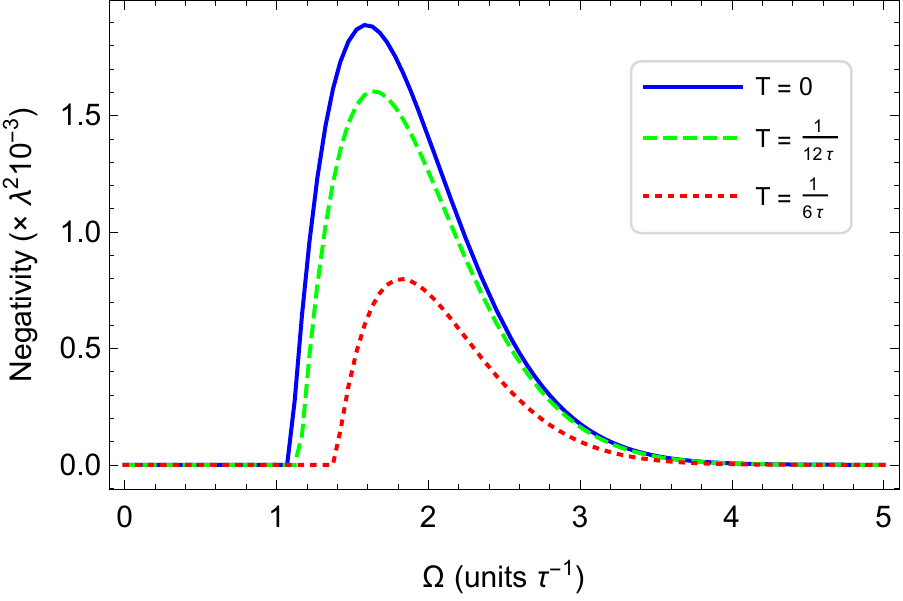}
    \caption{Negativity of identical detectors as a function of their energy gap, for different field temperatures $T$. The detectors are coupled to the field at the same time according to a Gaussian switching function of width $\uptau$, and they have Gaussian spatial profiles of width $\sigma=\uptau$, the centers of which are separated in space by $d=2\uptau$.}
    \label{fig:NegVsOmega_thermal}
\end{figure}

To conclude this section, let us briefly investigate how the negativity of the detectors varies with their energy gap $\Omega$. These results are summarized in Fig.~\ref{fig:NegVsOmega_thermal}. Notice that, for a given field temperature $T$, the detectors cannot become entangled if their energy gap is below some finite value $\Omega_\text{min}(T)$. We also notice that  $\Omega_\text{min}(T)$ is a monotonically increasing function of temperature. This tells us that if we have a way to control the energy gap of the detectors, then by measuring the amount of entanglement that this detector pair harvests from the field we have, in principle, a quantum thermometer capable of measuring the field temperature.

\subsection{Harvesting mutual information}
\label{eq:thermal:harvesting_mut_info}

Having shown that the amount of entanglement harvested by two Unruh-DeWitt detectors decreases with the temperature of the field with which they interact, we can ask what happens to other types of correlations. As mentioned above, the mutual information $I[\rhoab]$, defined in Eq.~\eqref{eq:mut_info}, quantifies the total correlations (quantum and classical) present between the two detectors. Using the time-evolved density matrix $\rhoab$ in Eq.~\eqref{eq:rhoab_thermal} for the two detectors, we find that $I[\rhoab]$ takes the form
\begin{align}
\label{eq:mut_info_2}
    I[\rhoab]=&
    \mathcal L_+\log(\mathcal L_+)+
    \mathcal L_-\log(\mathcal L_-)
    \\
    &-\mathcal L_\textsc{aa}\log(\mathcal L_\textsc{aa})
    -\mathcal L_\textsc{bb}\log(\mathcal L_\textsc{bb})
    +\mathcal O(\lambda^4)
    \notag,
\end{align}
where $\mathcal L_\pm$ is defined as
\begin{align}
    \mathcal L_\pm 
    =
    \frac{1}{2}
    \left(
    \mathcal L_\textsc{aa}+\mathcal L_\textsc{bb}
    \pm
    \sqrt{(\mathcal L_\textsc{aa}-\mathcal L_\textsc{bb})^2+4|\mathcal L_\textsc{ab}|^2}
    \right).
\end{align}

Although the general dependence of $I[\rhoab]$ on the temperature $\beta^{-1}$ is highly non-trivial, from Eq.~\eqref{eq:mut_info_2} it is straightforward to derive the asymptotic behaviour as $\beta^{-1}\rightarrow\infty$. Defining $\mathscr L_{\pm}:=\beta\mathcal L_{\pm}$ and $\mathscr L_{\nu\eta}:=\beta\mathcal{L}_{\nu\eta}$, we notice from Eq.~\eqref{eq:L_thermal} that $\mathscr L_{\pm}$ and $\mathscr L_{\nu\eta}$ are independent of $\beta$ in the limit $\beta^{-1}\rightarrow\infty$. Then from Eq.~\eqref{eq:mut_info_2} it is straightforward to show that in the $\beta^{-1}\rightarrow\infty$ limit the mutual information goes as
\begin{align}
    I[\rhoab]\sim
    \frac{1}{\beta}
    \big(&
    \mathscr L_{+}\log\mathscr L_{+}
    +
    \mathscr L_{-}\log\mathscr L_{-}
    \notag\\
    &-
    \mathscr L_{\textsc{aa}}\log\mathscr L_{\textsc{aa}}
    -
    \mathscr L_{\textsc{bb}}\log\mathscr L_{\textsc{bb}}
    \big).
\end{align}
Combining this with the fact that the mutual information is always non-negative, we conclude that in the large temperature limit (of course with a coupling constant small enough so that we are still within the perturbative regime) the total correlations that the detectors harvest from the field grow proportionally to the temperature $\beta^{-1}$.

\begin{figure}
    \centering
    \includegraphics[width=0.9\linewidth]{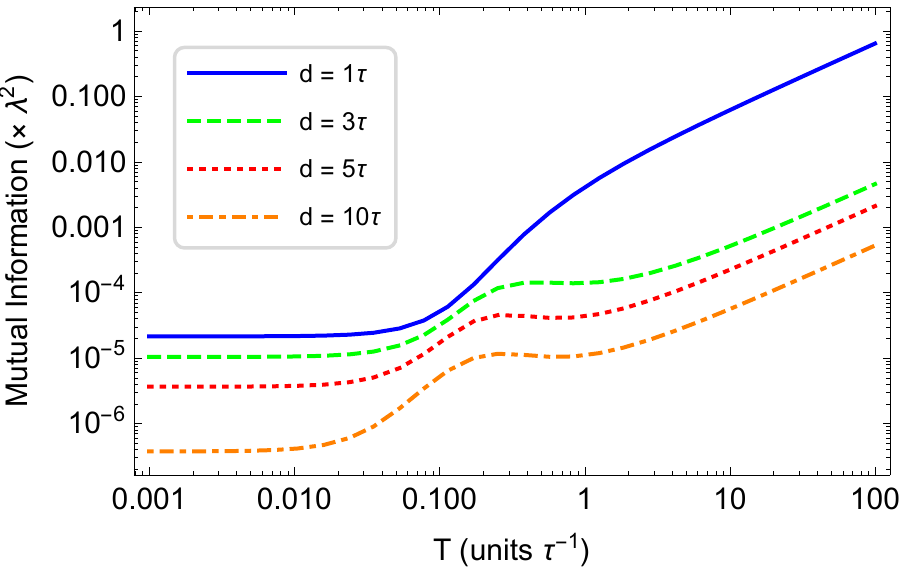}
    \caption{Mutual information of identical detectors as a function of field temperature, for different spatial separations $d$ of their centers of mass. The detectors are coupled to the field at the same time according to a Gaussian switching function of width $\uptau$, their spatial profiles are Gaussians of width $\sigma=\uptau$, and their energy gap is $\Omega=3/\uptau$.}
    \label{fig:MutInfoVsT_thermal}
\end{figure}

To see explicitly the dependence of $I[\rhoab]$ on the temperature, let us once again particularize to the case of identical detectors with Gaussian spatial smearings \eqref{eq:smearing_gaussian} and Gaussian switching functions \eqref{eq:switching_gaussian}. These results are plotted in Fig.~\ref{fig:MutInfoVsT_thermal}. We see that for low $T=\beta^{-1}$ the mutual information approaches a constant finite value, which corresponds to the correlations that the detectors would obtain if they interacted with the field vacuum. For intermediate field temperatures, we find that the mutual information has a non-trivial dependence on $T$, and in fact, unlike the negativity, $I[\rhoab]$ does not always increase with $T$. However, as we showed for the case of arbitrary detectors above, in the asymptotic limit $T\rightarrow\infty$ the mutual information is proportional to $T$. It should be emphasized that in a full, non-perturbative calculation, this upwards trend of $I[\rhoab]$ with temperature would not continue indefinitely for the simple reason that for a two qubit system the mutual information is bounded from above by $2\log 2$. Nevertheless it is interesting that, at least in the perturbative regime (i.e. if for a given temperature we consider a small enough coupling strength), the amount of entanglement harvested from the field by an Unruh-DeWitt detector pair is hindered by high field temperatures, whereas the total correlations in fact grow with $T$.

\section{Squeezed coherent field state}\label{sec:squeezed}

Again let us suppose that each Unruh-DeWitt detector is in its ground state, and that now the field is in an arbitrary, multimode, squeezed coherent state. The physical relevance of squeezed coherent states is that they are the most general set of states that saturate the Heisenberg uncertainty principle.  The most general multimode squeezed coherent state is given by $\ket{\alpha(\bm k),\zeta(\bm k,\bm k')}=\D\S\ket{0}$, where the \textit{displacement operato} $\D$ and the \textit{squeezing operator} $\S$ are unitary operators defined by~\cite{Loudon1987}
\begin{align}
    \label{eq:displacement_operator}
    \D&:=\exp\left[\int\d[3]{\bm k}\left(\alpha(\bm k)\ad k-\text{H.c.}\right)\right] ,
    \\
    \S&:=\exp\left[
    \frac{1}{2}\int\!\d[3]{\bm k}\!\!\int\!\d[3]{\bm k'}\left(\zeta^*(\bm k,\bm k')\a{\bm k}\a{\bm k'}-\text{H.c.}\right)
    \right]\!,
\end{align}
We call the complex valued distributions $\alpha(\bm k)$ and $\zeta(\bm k,\bm k')$  respectively the \textit{coherent amplitude} and \textit{squeezing amplitude} of the state $\ket{\alpha(\bm k),\zeta(\bm k,\bm k')}$. Through the integrals in the definitions of $\D$ and $\S$, these distributions generalize the familiar notion of a squeezed coherent state of a single harmonic oscillator to the case where we have an uncountably infinite number of field mode oscillators that can be pairwise two-mode squeezed with each other. 

In order to calculate the one and two-point functions of the field in a squeezed coherent state, we will make use of the identities governing the action of $\D$ and $\S$ on the creation and annihilation operators. Namely, by using the canonical commutation relations and the Baker-Campbell-Hausdorff lemma it is straightforward to show that
\begin{align}
    \D^\dagger\a k\D&=
    \a k+\alpha(\bm k)\openone.
\end{align}
On the other hand, we are not aware of a similarly convenient closed-form expression for $\S^\dagger\a k\S$ in the case of an arbitrary, continuous, multimode squeezing. However, since $\S$ is the exponential of terms quadratic in $\a k$ and $\ad k$, by expanding out the exponentials in $\S^\dagger\a k\S$ it is not difficult to prove that this expression takes the form of a linear superposition of $\a k$ and $\ad k$ operators, i.e.
\begin{equation}
\label{eq:SaS}
    \S^\dagger\a k\S 
    =
    \int\d[3]{\bm k'}
    \left[K_1(\bm k,\bm k')\a{k'}
    +
    K_2(\bm k,\bm k')\ad{k'}
    \right],
\end{equation}
for some bi-distributions $K_1$ and $K_2$. In particular this implies that
\begin{align}
    &\bra{\alpha(\bm k),\zeta(\bm k,\bm k')}\a {k''}\ket{\alpha(\bm k),\zeta(\bm k,\bm k')}\notag\\
    =
    \,&\bra{0}\S^\dagger[\a {k''}+\alpha(\bm k'')]\S\ket{0}\notag\\
    =
    \,&\alpha(\bm k''),
\end{align}
and hence, using the mode expansion~\eqref{eq:field} of the field operator, the one-point function~\eqref{eq:v} of the field in the state $\ket{\alpha(\bm k),\zeta(\bm k,\bm k')}$ is 
\begin{align}
\label{eq:v_sqeezed_coherent}
    v(\bm x,t)=
    \int\frac{\d[n]{\bm k}}{\sqrt{2(2\pi)^n\om k}}
    \left(
    \alpha(\bm k)e^{-\ii(\om k t-\bm k\cdot \bm x)}
    +
    \text{c.c}
    \right).
\end{align}
Thus we see that the one-point function is independent of the squeezing amplitude $\zeta(\bm k,\bm k')$. Similarly we can show that the two-point function~\eqref{eq:w} in the state $\ket{\alpha(\bm k),\zeta(\bm k,\bm k')}$ is of the form
\begin{align}
\label{eq:w_general_sqeezed_coherent}
    w(\bm x, t,\bm x',t')
    =
    w^\text{ind}(\bm x, t,\bm x',t')
    +
    w^\text{coh}(\bm x, t,\bm x',t'),
\end{align}
where $w^\text{ind}$ is independent of the coherent amplitude $\alpha(\bm k)$, where $w^\text{coh}$ is given by a product of one-point functions:
\begin{equation}
\label{eq:wcoh}
    w^\text{coh}(\bm x,t,\bm x',t')
    =
    v(\bm x,t)v(\bm x',t'),
\end{equation}
and vanishes if $\alpha(\bm k)= 0$ for all $\bm k$.

Even without calculating the $\alpha(\bm k)$-independent contribution $w^\text{ind}$ to the two-point function, we can see that it is the product of two one-point functions. In \cite{Simidzija2017c} it was shown that when this is the case, then the $\alpha(\bm k)$-dependent contributions of $\rhoab$ arising from the one-point function exactly cancel the contributions from the two-point function, so that the eigenvalues of $\rhoab$ and $\rhoabpt$ --- and therefore the negativity $\mathcal N[\rhoab]$ as well --- are completely independent of $\alpha(\bm k)$. This result was used in \cite{Simidzija2017c} to prove that the entanglement harvested by an Unruh-DeWitt detector pair is independent of the coherent amplitude of a (non-squeezed) coherent state. Since this is a general consequence of the special relationship between the $\alpha(\bm k)$-dependent parts of the one and two-point functions, we conclude that this result is true even in the presence of squeezing. Namely, to $\mathcal O(\lambda^2)$, the negativity of a detector pair interacting with a general squeezed coherent state $\ket{\alpha(\bm k),\zeta(\bm k,\bm k')}$ is independent of the coherent amplitude distribution $\alpha(\bm k)$. In other words, entanglement harvesting from a squeezed coherent state is insensitive to the coherent amplitude.

Therefore, since we are interested in studying the entanglement harvested by the detector pair from a general squeezed coherent state, we can, without loss of generality, restrict our attention only to squeezed vacuum states (i.e. we can make $\alpha(\bm k)$ identically zero). Additionally, for mathematical simplicity---i.e. in order to obtain an explicit expression for $\S^\dagger\a k\S$ in Eq.~\eqref{eq:SaS}---from here on we will consider only squeezed coherent states in which the squeezing is not ``mixed" between modes, i.e. such that the squeezing amplitude is of the form $\zeta(\bm k,\bm k')=\zeta(\bm k)\delta(\bm k-\bm k')$. In this case we find that $\S$ simplifies to 
\begin{equation}
    \S=\exp\left[
    \frac{1}{2}\int\d[3]{\bm k}\left(\zeta^*(\bm k)\hat a_{\bm k}^2-\text{H.c.}\right)
    \right],
\end{equation}
and that $\S^\dagger\a k\S$ can be conveniently expressed as
\begin{align}
    \S^\dagger\a k\S&=
    \cosh[r(\bm k)]\a k-e^{\ii\theta(\bm k)}\sinh[r(\bm k)]\ad k,
\end{align}
where we have written $\zeta(\bm k)=r(\bm k)e^{\ii\theta(\bm k)}$ in polar form. The two-point function~\eqref{eq:w} of the state $\S\ket{0}$, with $\S$ in the above form, can be written as
\begin{align}
    \label{eq:w_squeezed}
    w(\bm x, t,\bm x',t')
    =
    w^\text{vac}(\bm x, t,\bm x',t')
    +
    w^\text{sq}(\bm x, t,\bm x',t'),
\end{align}
where $w^\text{vac}$ is the vacuum two-point function given in Eq.~\eqref{eq:wvac}, while $w^\text{sq}$ is the contribution that depends on $\zeta(\bm k)$ and vanishes if $\zeta(\bm k)=0$ for all $\bm k$. Explicitly $w^\text{sq}(\bm x,t,\bm x',t')$ is given by
\begin{align}
\label{eq:wsq}
    &w^\text{sq}(\bm x,t,\bm x', t')
    =
    \int\frac{\d[n]{\bm k}}{2(2\pi)^n\om k}
    \sinh[r(\bm k)]
    \\
    &\hspace{0.5cm}\times
    \Big(
    \!\!-e^{\ii\theta(\bm k)}\cosh[r(\bm k)]
    e^{-\ii\om k(t+t')}e^{\ii\bm k\cdot(\bm x+\bm x'}
    \notag\\
    &\hspace{1cm}
    +\sinh[r(\bm k)]
    e^{-\ii\om k(t-t')}e^{\ii\bm k\cdot(\bm x-\bm x'}
    \Big)
    +\text{c.c},
\end{align}
Notice that, unlike Eq.~\eqref{eq:thermal_wightman} for a thermal field state, the two-point function for a squeezed coherent state is not invariant with respect to spacetime translations. As we will see, a physical consequence of this is that the negativity harvested by a pair of UDW detectors from a squeezed coherent state depends not only on the spacetime interval between the detectors, but also on where in the spacetime they are centered. 

With the expression~\eqref{eq:w_squeezed} for the two-point function of a squeezed vacuum field state, and with the vanishing one-point function~\eqref{eq:v_sqeezed_coherent}, we can proceed to calculate the evolved state $\rhoab$ of the two UDW detectors following their interactions with this field. From~\eqref{eq:rhoab^2} we obtain
\begin{equation}
	\label{eq:rhoab_squeezed}
    \rhoab=
    \begin{pmatrix}
	1-\mathcal{L}_\textsc{aa}[\zeta]-\mathcal{L}_\textsc{bb}[\zeta]& 0 & 0 & \mathcal{M}^*[\zeta] \\
	0 & \mathcal{L}_\textsc{bb}[\zeta] & \mathcal{L}_\textsc{ab}^*[\zeta] & 0 \\
	0 & \mathcal{L}_\textsc{ab}[\zeta] & \mathcal{L}_\textsc{aa}[\zeta] & 0 \\
	\mathcal{M}[\zeta] & 0 & 0 & 0
	\end{pmatrix},
\end{equation}
to second order in the coupling strength $\lambda$, and where we work in the basis $\{    \ket{g_\textsc{a}}\ket{g_\textsc{b}},
\ket{g_\textsc{a}}\ket{e_\textsc{b}},
\ket{e_\textsc{a}}\ket{g_\textsc{b}},
\ket{e_\textsc{a}}\ket{e_\textsc{b}}\}$. The matrix terms $\mathcal{L}_{\nu\eta}[\zeta]$ and $\mathcal M[\zeta]$ are now functionals of the squeezing distribution $\zeta(\bm k)$, and they take the forms
\begin{align}
    \mathcal L_{\nu\eta}[\zeta]&=
    \mathcal L_{\nu\eta}^\text{vac}
    +
    \mathcal L_{\nu\eta}^\text{sq}[\zeta],
    \\
    \mathcal M[\zeta]&=\mathcal M^\text{vac}+\mathcal M^\text{sq}[\zeta].
\end{align}
As before, the vacuum terms $\mathcal L_{\nu\eta}^\text{vac}$ and $\mathcal M^\text{vac}$ are given by Eqs.~\eqref{eq:Lvac} and \eqref{eq:Mvac}, while the $\zeta(\bm k)$ dependent terms read
\begin{widetext}
\begin{align}
    \label{eq:L_squeezed_general}
    \mathcal L_{\nu\eta}^\text{sq}[\zeta]
    &=
    \pi\lambda_\nu\lambda_\eta
    \int\frac{\d[3]{\bm k}}{\om k}
    \Big(
    \sinh^2[r(\bm k)]
    \bar F_\nu(\bm k)
    \bar F_\eta^*(\bm k)
    \bar \chi_\nu^*(\om k-\Omega_\nu)
    \bar \chi_\eta(\om k-\Omega_\eta)
    e^{\ii\bm k\cdot(\bm x_\nu-\bm x_\eta)}
    \\
    &\hspace{3cm}
    +
    \sinh^2[r(\bm k)]
    \bar F_\nu^*(\bm k)
    \bar F_\eta(\bm k)
    \bar \chi_\nu(\om k+\Omega_\nu)
    \bar \chi_\eta^*(\om k+\Omega_\eta)
    e^{-\ii\bm k\cdot(\bm x_\nu-\bm x_\eta)}
    \notag
    \\
    &\hspace{3cm}
    -
    e^{-\ii\theta(\bm k)}
    \sinh[r(\bm k)]\cosh[r(\bm k)]
    \bar F_\nu^*(\bm k)
    \bar F_\eta^*(\bm k)
    \bar \chi_\nu(\om k+\Omega_\nu)
    \bar \chi_\eta(\om k-\Omega_\eta)
    e^{-\ii\bm k\cdot(\bm x_\nu+\bm x_\eta)}
    \notag
    \\
    &\hspace{3cm}
    -
    e^{\ii\theta(\bm k)}
    \sinh[r(\bm k)]\cosh[r(\bm k)]
    \bar F_\nu(\bm k)
    \bar F_\eta(\bm k)
    \bar \chi_\nu^*(\om k-\Omega_\nu)
    \bar \chi_\eta^*(\om k+\Omega_\eta)
    e^{\ii\bm k\cdot(\bm x_\nu+\bm x_\eta)}
    \Big),
    \notag
    \\
    \label{eq:M_squeezed_general}
    \mathcal M^\text{sq}[\zeta]
    &=
    2\pi\lambda_\textsc{a}\lambda_\textsc{b}
    \int\frac{\d[3]{\bm k}}{\om k}
    \Big(
    e^{-\ii\theta(\bm k)}
    \sinh[r(\bm k)]\cosh[r(\bm k)]
    \bar F_\textsc{a}^*(\bm k)
    \bar F_\textsc{b}^*(\bm k)
    \bar \chi_\textsc{a}(\om k+\Omega_\textsc{a})
    \bar \chi_\textsc{b}(\om k+\Omega_\textsc{b})
    e^{-\ii\bm k\cdot(\bm x_\textsc{a}+\bm x_\textsc{b})}
    \\
    &\hspace{3cm}
    +
    e^{\ii\theta(\bm k)}
    \sinh[r(\bm k)]\cosh[r(\bm k)]
    \bar F_\textsc{a}(\bm k)
    \bar F_\textsc{b}(\bm k)
    \bar \chi_\textsc{a}^*(\om k-\Omega_\textsc{a})
    \bar \chi_\textsc{b}^*(\om k-\Omega_\textsc{b})
    e^{\ii\bm k\cdot(\bm x_\textsc{a}+\bm x_\textsc{b})}
    \notag
    \\
    &\hspace{3cm}
    -
    \sinh^2[r(\bm k)]
    \bar F_\textsc{a}^*(\bm k)
    \bar F_\textsc{b}(\bm k)
    \bar \chi_\textsc{a}(\om k+\Omega_\textsc{a})
    \bar 
    \chi_\textsc{b}^*(\om k-\Omega_\textsc{b})
    e^{-\ii\bm k\cdot(\bm x_\textsc{a}-\bm x_\textsc{b})}
    \notag
    \\
    &\hspace{3cm}
    -
    \sinh^2[r(\bm k)]
    \bar F_\textsc{a}(\bm k)
    \bar F_\textsc{b}^*(\bm k)
    \bar \chi_\textsc{a}^*(\om k-\Omega_\textsc{a})
    \bar \chi_\textsc{b}(\om k+\Omega_\textsc{b})
    e^{\ii\bm k\cdot(\bm x_\textsc{a}-\bm x_\textsc{b})}
    \Big).
    \notag
\end{align}
\end{widetext}

\subsection{Harvesting entanglement}

In order to study the dependence of field squeezing on the ability of detectors to harvest entanglement, let us once again particularize to the case of a massless field and identical UDW detectors with Gaussian spatial profiles of width $\sigma$, given by Eq.~\eqref{eq:smearing_gaussian}, and Gaussian temporal switching functions of width $\uptau$, as in Eq.~\eqref{eq:switching_gaussian}. Then the matrix elements $\mathcal L_{\nu\eta}^\text{sq}[\zeta]$ and $\mathcal M^\text{sq}[\zeta]$ given by Eqs.~\eqref{eq:L_squeezed_general} and~\eqref{eq:M_squeezed_general} become
\begin{widetext}
\begin{align}
    \label{eq:L_squeezed_gaussian}
    \mathcal L_{\nu\eta}^\text{sq}[\zeta]
    &=
    \frac{\tilde\lambda^2 e^{-\frac{1}{2}\tilde\Omega^2}}{16\pi^2}
    \int\frac{\d[3]{\tilde{\bm k}}}{|\tilde{\bm k}|}
    e^{-\frac{1}{2}|\tilde{\bm k}|^2(1+\tilde\sigma^2)}
    \Big(
    \sinh^2[r(\bm k)]
    e^{|\tilde{\bm k}|\tilde\Omega}
    e^{-\ii|\tilde{\bm k}|(\tilde t_\nu-\tilde t_\eta)}
    e^{\ii\tilde{\bm k}\cdot(\tilde{\bm x}_\nu-\tilde{\bm x}_\eta)}
    \\
    &\hspace{5cm}
    +
    \sinh^2[r(\bm k)]
    e^{-|\tilde{\bm k}|\tilde\Omega}
    e^{\ii|\tilde{\bm k}|(\tilde t_\nu-\tilde t_\eta)}
    e^{-\ii\tilde{\bm k}\cdot(\tilde{\bm x}_\nu-\tilde{\bm x}_\eta)}
    \notag
    \\
    &\hspace{5cm}
    -
    e^{-\ii\theta(\bm k)}
    \sinh[r(\bm k)]\cosh[r(\bm k)]
    e^{\ii|\tilde{\bm k}|(\tilde t_\nu+\tilde t_\eta)}
    e^{-\ii\tilde{\bm k}\cdot(\tilde{\bm x}_\nu+\tilde{\bm x}_\eta)}
    \notag
    \\
    &\hspace{5cm}
    -
    e^{\ii\theta(\bm k)}
    \sinh[r(\bm k)]\cosh[r(\bm k)]
    e^{-\ii|\tilde{\bm k}|(\tilde t_\nu+\tilde t_\eta)}
    e^{\ii\tilde{\bm k}\cdot(\tilde{\bm x}_\nu+\tilde{\bm x}_\eta)}
    \notag
    \Big),
    \notag
    \\
    \label{eq:M_squeezed_gaussian}
    \mathcal M^\text{sq}[\zeta]
    &=
    -\frac{\tilde\lambda^2 e^{-\frac{1}{2}\tilde\Omega^2}}{16\pi^2}
    \int\frac{\d[3]{\tilde{\bm k}}}{|\tilde{\bm k}|}
    e^{-\frac{1}{2}|\tilde{\bm k}|^2(1+\tilde\sigma^2)}
    \Big(
    \sinh^2[r(\bm k)]
    e^{-\ii|\tilde{\bm k}|(\tilde t_\textsc{a}-\tilde t_\textsc{b})}
    e^{\ii\tilde{\bm k}\cdot(\tilde{\bm x}_\textsc{a}-\tilde{\bm x}_\textsc{b})}
    \\
    &\hspace{5cm}
    +
    \sinh^2[r(\bm k)]
    e^{\ii|\tilde{\bm k}|(\tilde t_\textsc{a}-\tilde t_\textsc{b})}
    e^{-\ii\tilde{\bm k}\cdot(\tilde{\bm x}_\textsc{a}-\tilde{\bm x}_\textsc{b})}
    \notag
    \\
    &\hspace{5cm}
    -
    e^{-\ii\theta(\bm k)}
    \sinh[r(\bm k)]\cosh[r(\bm k)]
    e^{-|\tilde{\bm k}|\tilde\Omega}
    e^{\ii|\tilde{\bm k}|(\tilde t_\textsc{a}+\tilde t_\textsc{b})}
    e^{-\ii\tilde{\bm k}\cdot(\tilde{\bm x}_\textsc{a}+\tilde{\bm x}_\textsc{b})}
    \notag
    \\
    &\hspace{5cm}
    -
    e^{\ii\theta(\bm k)}
    \sinh[r(\bm k)]\cosh[r(\bm k)]
    e^{|\tilde{\bm k}|\tilde\Omega}
    e^{-\ii|\tilde{\bm k}|(\tilde t_\textsc{a}+\tilde t_\textsc{b})}
    e^{\ii\tilde{\bm k}\cdot(\tilde{\bm x}_\textsc{a}+\tilde{\bm x}_\textsc{b})}
    \notag
    \Big),
    \notag
\end{align}
\end{widetext}
where, as before, we denote by a tilde any quantity referred to the scale $\uptau$ (e.g., $\tilde\Omega=\Omega\uptau$, $\tilde \sigma =\sigma/\uptau$, etc.). With these explicit expressions for the matrix elements of $\rhoab$ at hand, we can now readily compute the negativity $\mathcal N=\max\left(0,|\mathcal M[\zeta]|-\mathcal L_{\nu\nu}[\zeta]\right)$, and thus quantify the amount of entanglement that the two detectors harvest from the field. 

\subsubsection{Uniform squeezing}
\label{sec:uniform_squeezing}

Let us begin by considering the simplest possible type of squeezing: that in which all field modes are squeezed equally. To that end we take $\zeta(\bm k)=r$, where we also assume that $r$ is real and positive. (We will shortly see what the effect is of $r$ having a complex phase.)

\begin{figure}
    \centering
    \includegraphics[width=0.9\linewidth]{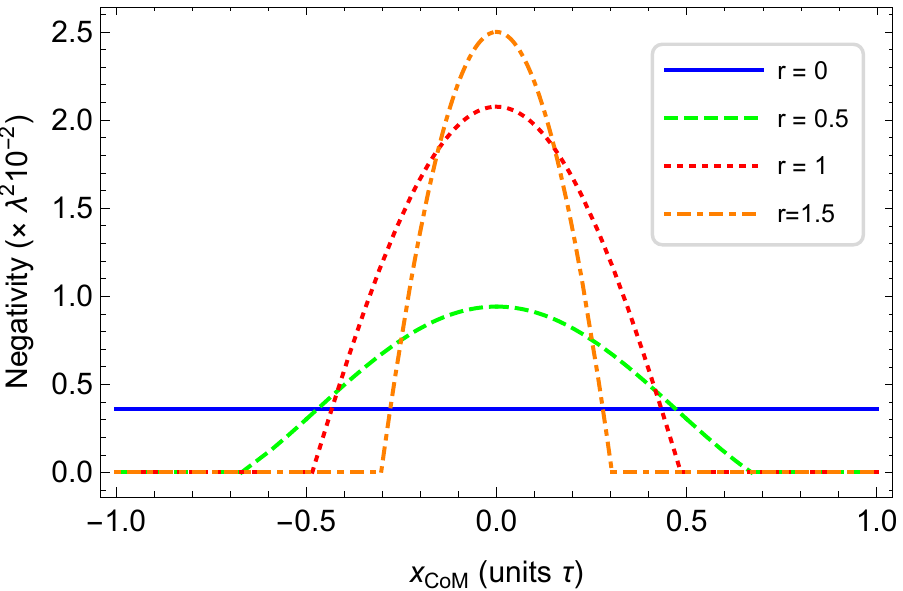}
    \caption{Negativity of identical detectors as a function of their center of mass position, for different values of the squeezing parameter $r=|\zeta(\bm k)|$. Here the squeezing is uniform across all field modes. The detectors are coupled to the field through Gaussian switching functions of width $\uptau$ centered at $t=0$, and their energy gaps are $\Omega=\uptau^{-1}$. The detectors are centered at $(x_\textsc{com}\pm \uptau,0,0)$ and have Gaussian spatial profiles of width $\sigma=\uptau$.}
    \label{fig:neg_vs_xcom}
\end{figure}

In Fig.~\ref{fig:neg_vs_xcom}, for different values of $r$, we plot the negativity of the detectors following their interactions with the field as a function of their joint center of mass. We see that---as we anticipated already from the two-point function---a squeezed field state is in general not translationally invariant, and as such the entanglement harvesting ability of a pair of detectors from such a state is not translationally invariant either. In particular we find that if the detectors' center of mass is near the spatial origin of the coordinate system, then the detectors can harvest more entanglement from a uniformly squeezed field state than from the vacuum. On the other hand if the detectors are far enough away from the origin, then, regardless of the amount of squeezing, they are unable to extract entanglement. The proximity to the origin that is necessary for squeezing to be beneficial for entanglement harvesting is dictated by the amount of squeezing $r$: for a highly squeezed field state the detectors can harvest a lot more entanglement, but they have to be highly centered near the origin; for a less squeezed state the improvement in harvesting is not as noticeable, but the detectors do not need to be so precisely centered.

Let us now attempt to better understand the non-translation-invariance of squeezed field states in general, and in particular the consequences of this for entanglement harvesting from these states. Concretely, with regards to the plots in Fig.~\ref{fig:neg_vs_xcom}, it is natural to ask why is the spatial origin of our chosen coordinate system the preferred location of UDW detectors that hope to harvest entanglement? First, let us note once again that, as can be seen in Fig~\ref{fig:neg_vs_xcom}, in the absence of squeezing the translation-invariance of entanglement harvesting is restored. Therefore, the picking out of a preferred point in space near which entanglement harvesting is maximized (in this case the origin of the coordinate system) must be a direct consequence of the squeezing amplitude $\zeta(\bm k)$ that we choose for the field. In fact, we notice that the Fourier transform of the uniform amplitude $\zeta(\bm k)=r$ is proportional to $\delta(\bm x)$, and therefore the origin $\bm x=0$ is clearly a special point in this case. As we will now show, this relationship between the Fourier transform of the squeezing amplitude and the preferred location of detectors trying to harvest entanglement is valid in general.

To that end, let us consider an arbitrary squeezing amplitude $\zeta(\bm k)$. With this choice of squeezing, there will be some preferred points in space near which it is easier for detectors to harvest entanglement, and others near which it is more difficult. Suppose now that we change the squeezing by a local phase $\zeta(\bm k) \rightarrow \zeta'(\bm k)=e^{\ii\bm k\cdot\bm x_0} \zeta(\bm k)$. How do the positions of the preferred points change?

To answer this question, let us recall from Eq.~\eqref{eq:rhoab1} that the state $\rhoab$ of the two detectors following their interactions with a squeezed field state with amplitude $\zeta'$ is given by
\begin{equation}
    \rhoab=\tr_\phi\left[\hat U'\left(\rhoa\otimes\rhob\otimes\hat S_{\zeta'}^\dagger\ket{0}\bra{0}\hat S_{\zeta'}\right)\hat U'^\dagger\right],
\end{equation}
where $\hat U'$ is the time-evolution unitary
\begin{align}
	\hat{U'}
	=
	\mathcal{T}\exp
	\Big[
	&-\ii\!\int\!\dif t 
	\sum_\nu
	\lambda_\nu \chi_\nu(t) \hat{\mu}_\nu(t)
	\\
	&\times
	\int \!\d[n]{\bm{x}} F_\nu(\bm{x}-\bm{x}_\nu) \hat{\phi}(\bm{x},t)
	\Big].
	\notag
\end{align}
Now let us define the field \textit{momentum operator} to be $\hat{\bm P}:=\int\d[3]{\bm k} \bm k\ad k\a k$. Then, using the fact that
\begin{equation}
    \label{eq:translation_of_a}
    e^{\ii\hat{\bm P}\cdot \bm x_0}\a k e^{-\ii\hat{\bm P}\cdot \bm x_0}=\a k e^{\ii\bm k\cdot\bm x_0},
\end{equation}
we find that we can write $\hat S_{\zeta'}=e^{-\ii\hat{\bm P}\cdot \bm x_0/2}\hat S_{\zeta}e^{\ii\hat{\bm P}\cdot \bm x_0/2}$. Making use of the cyclicity of the partial trace with respect to the subsystem being traced over, we find that $\rhoab$ can be expressed as
\begin{equation}
    \rhoab=\tr_\phi\left[\hat U\left(\rhoa\otimes\rhob\otimes\hat S_{\zeta}^\dagger\ket{0}\bra{0}\hat S_{\zeta}\right)\hat U^\dagger\right],
\end{equation}
where $\hat U:=e^{\ii\hat{\bm P}\cdot \bm x_0/2}\hat U'e^{-\ii\hat{\bm P}\cdot \bm x_0/2}$. Using~\eqref{eq:translation_of_a} we readily obtain
\begin{align}
	\hat{U}
	=
	\mathcal{T}\exp
	\Big[
	&-\ii\!\int\!\!\dif t\! 
	\sum_\nu
	\lambda_\nu \chi_\nu(t) \hat{\mu}_\nu(t)
	\\
	&\times
	\int \!\d[n]{\bm{x}} F_\nu\left(\bm{x}-\bm{x}_\nu-\frac{\bm x_0}{2}\right) \hat{\phi}(\bm{x},t)
	\Big].
	\notag
\end{align}
Hence changing the field's squeezing amplitude by a local phase $\zeta \rightarrow e^{\ii\bm k\cdot\bm x_0} \zeta$ is equivalent to shifting the detectors in space by an amount $\bm x_0/2$. In other words, a local phase change of the squeezing amplitude effects a translation of the points in space near which it is easier for the detectors to harvest entanglement. However, such a local phase change of $\zeta$ also effects a translation of its Fourier transform: namely $\bar\zeta(\bm x)\rightarrow \bar \zeta(\bm x-\bm x_0)$. Note that the discrepancy by a factor of 2 between the amount that the preferred points are translated ($\bm x_0/2$) and the amount that the Fourier transform $\bar\zeta$ is shifted by ($\bm x_0$) can be removed by choosing a different convention for the exponent in the definition~\eqref{eq:FT} of a Fourier transform. Therefore we conclude that (up to a potential re-scaling) the Fourier transform of the field's squeezing amplitude $\zeta$ directly tells us where in space the UDW detectors should be centered if they want to harvest more entanglement from the squeezed field state. These preferred locations are commensurate with where the fluctuations of the field amplitude, and the stress energy density, are localized in space.

\begin{figure}
    \centering
    \subfigure{
		\includegraphics[width=0.45\textwidth]{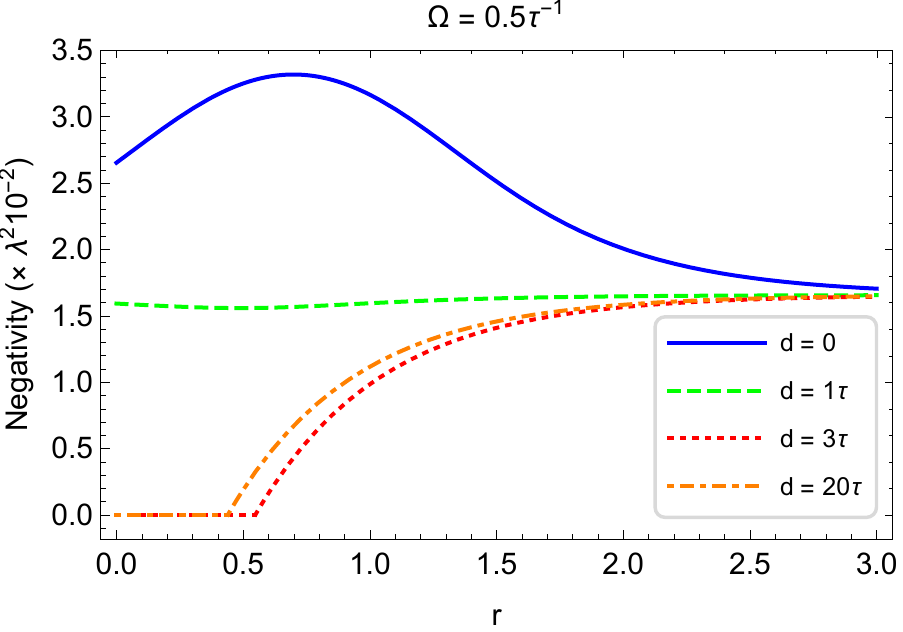}}
	\subfigure{
		\includegraphics[width=0.45\textwidth]{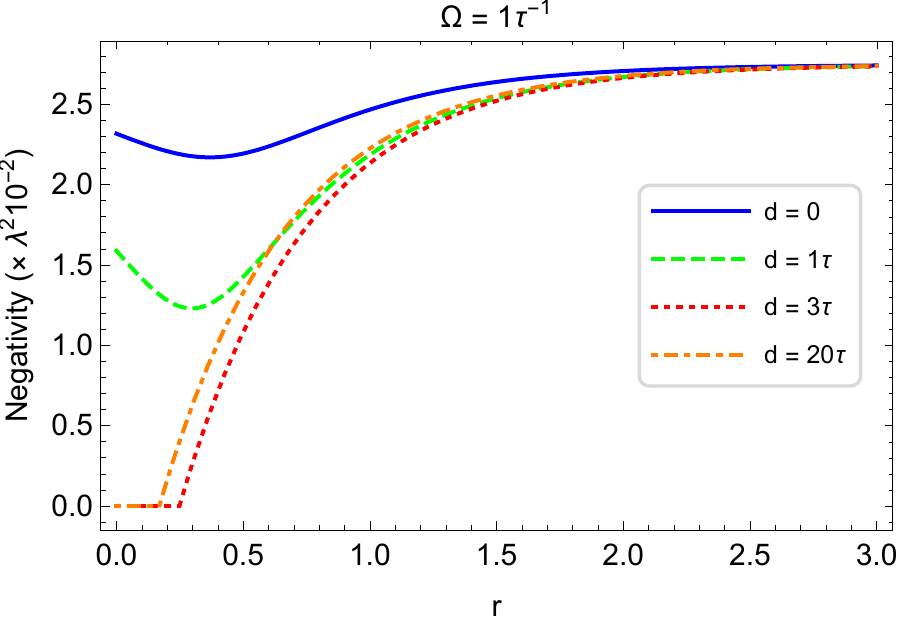}}
	\caption{Negativity of identical detectors as a function of the squeezing parameter $r=|\zeta(\bm k)|$, for different values of their spatial separation $d$ and energy gaps $\Omega$. Here the squeezing is uniform across all field modes. The detectors are coupled to the field through Gaussian switching functions of width $\uptau$ centered at $t=0$; they are centered at $(\pm d/2,0,0)$ and have Gaussian spatial profiles of width $\sigma=\uptau$.}
	\label{fig:neg_vs_r}
\end{figure}

Having expounded the dependence of the detectors' center of mass on their ability to harvest entanglement from a squeezed field state, and having related this to the local phase of the squeezing amplitude, let us now turn to the question of how the \textit{magnitude} of the squeezing amplitude affects the detector's abilities to harvest entanglement. 

In Fig.~\ref{fig:neg_vs_r} we plot the negativity of a UDW detector pair as a function of $\zeta(\bm k)=r$, which we once again assume to be uniform across all field modes. We notice several interesting features from these plots.

Interestingly, high squeezing can remove the dependence of entanglement harvesting on the distance between the detectors. Indeed, we find that while at low squeezing amplitude the amount of entanglement that the detectors can harvest depends on their spatial separation $d:=|\bm x_\textsc{a}-\bm x_\textsc{b}|$, at high squeezing this is not the case. In other words, in the limit of large uniform squeezing of the field, a detector pair separated by a large spatial distance will harvest the same amount of entanglement as if they were at the same location in space. A similar effect of removal of the distance scale in a setup where vacuum entanglement is relevant was seen in \cite{Hotta2014} where quantum energy teleportation could be made independent of separation between sender and receiver if one uses squeezed field states.

Furthermore, from Fig.~\ref{fig:neg_vs_r}, we find that the amount of entanglement that the detectors harvest is also independent of the squeezing parameter $\zeta(\bm k)=r$ in the limit as $r\rightarrow\infty$. Hence although squeezing the field modes often increases the amount of harvestable entanglement from that allowed by the field vacuum, this trend of increasing negativity does not continue indefinitely, but rather plateaus to a constant asymptotic value at large $r$.

\subsubsection{Bandlimited squeezing}
\label{sec:bandlimited_squeezing}

To an experimentalist looking to make an entanglement harvesting measurement in the lab, perhaps the most interesting results of the previous section are that i) the amount of entanglement harvested by a pair of UDW detectors from a highly (uniformly) squeezed field state is \textit{independent} of the spatial separation of the detectors, and ii) if the detectors are centered near the ``preferred" locations in space (as determined by the Fourier transform of the squeezing function $\zeta(\bm k)$), then the amount of entanglement that they harvest could be much higher than in the case of a vacuum field state.

However such an experimentalist would be quick to note that there is an obvious difficulty with attempting to translate the theoretical results of the previous section into an actual experiment in the lab. Namely, in the previous section we assumed the field to be uniformly squeezed across all field modes, while squeezed states in experimental quantum optics~\cite{Bachor2004} and superconducting setups~\cite{Zagoskin2008} are generally bandlimited to a very narrow range of field modes. We expect that in this case, where only a narrow frequency range of modes are squeezed, the field state will behave more similarly to the vacuum state, in which case squeezing might not give much of an advantage in terms of entanglement harvesting. The key question is then: what range of field modes must be squeezed in order to produce a significant entanglement harvesting advantage over the vacuum state?

To answer this question, let us now assume that only the field modes \textit{near} some momentum $\bm k$ are uniformly squeezed, while all other modes are in their vacuum states. More precisely, we set
\begin{equation}
    \label{eq:bandlimited_squeezing}
    \zeta(\bm k') =
    \begin{cases}
        r & \text{if }|k'_i-k_i|<\frac{\epsilon}{2} \text{ for } i\in\{x,y,z\}\\
        0 & \text{otherwise}
    \end{cases},
\end{equation}
where $\bm k' = (k'_x,k'_y,k'_z)$, $\bm k = (k_{x},k_{y},k_{z})$, and $\epsilon$ parametrizes the bandwidth of the squeezing. With this choice of squeezing amplitude, and assuming again that the spatial and temporal profiles of the detectors are Gaussians given by Eqs.~\eqref{eq:smearing_gaussian} and \eqref{eq:switching_gaussian}, the matrix elements $\mathcal L_{\nu\eta}^\text{sq}[\zeta]$ and $\mathcal M^\text{sq}[\zeta]$ of the evolved two detector density matrix $\rhoab$ are again given by the expressions in Eqs.~\eqref{eq:L_squeezed_gaussian} and \eqref{eq:M_squeezed_gaussian}, except that now the limits of momentum space integration are such that $|k'_i-k_{i}|<\epsilon/2$. With the use of these expressions we can compute the negativity $\mathcal N=\max\left(0,|\mathcal M[\zeta]|-\mathcal L_{\nu\nu}[\zeta]\right)$, and thus observe how the amount of entanglement that the detectors can harvest depends on the bandwidth $\epsilon$ of the field's squeezing amplitude.

However before showing plots of $\mathcal N$ versus $\epsilon$, since we are in this section trying to upgrade our theoretical findings to the realm of what is experimentally feasible, it is important that we also discuss what values of squeezing amplitude $r$ we can expect to obtain in our bandlimited frequency range. As far as we are aware, the highest experimentally attained squeezed state of the electromagnetic field resulted in a squeezed quadrature noise reduction of 15 dB below the vacuum level~\cite{Henning2016}. Using the conversion formula~\cite{Lvovsky2015}
\begin{equation}
    \Delta\text{Noise} \text{ (in dB)}=
    10 \log_{10}\left(2\langle\Delta \hat X^2\rangle\right),
\end{equation}
between the reduction in noise of the squeezed quadrature $\hat X$ and the variance $\langle\Delta \hat X^2\rangle:=\langle\hat X^2\rangle-\langle \hat X\rangle^2$ of that quadrature in the squeezed state $\ket{\zeta(\bm k)}$, as well as the expression
\begin{equation}
    \langle\Delta \hat X^2\rangle
    =
    \frac{1}{2}e^{-2r},
\end{equation}
between $\langle\Delta \hat X^2\rangle$ and $r$, we find that
\begin{equation}
    \Delta\text{Noise} \text{ (in dB)} =
    -20\log_{10}(e)r.
\end{equation}
Hence a noise reduction of 15 dB corresponds to a squeezing amplitude of $r\approx 1.7$. To be on the safe side with respect to experimental feasibility, we will for the below discussion set $r=1$ (corresponding to $\sim 8.7$ dB).

\begin{figure}
    \centering
    \subfigure{
		\includegraphics[width=0.45\textwidth]{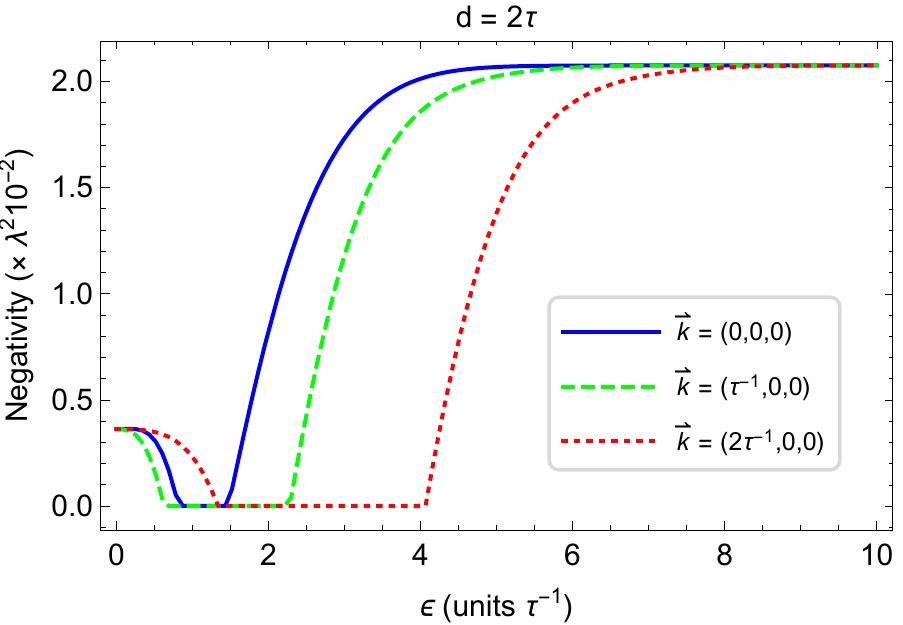}}
	\subfigure{
		\includegraphics[width=0.45\textwidth]{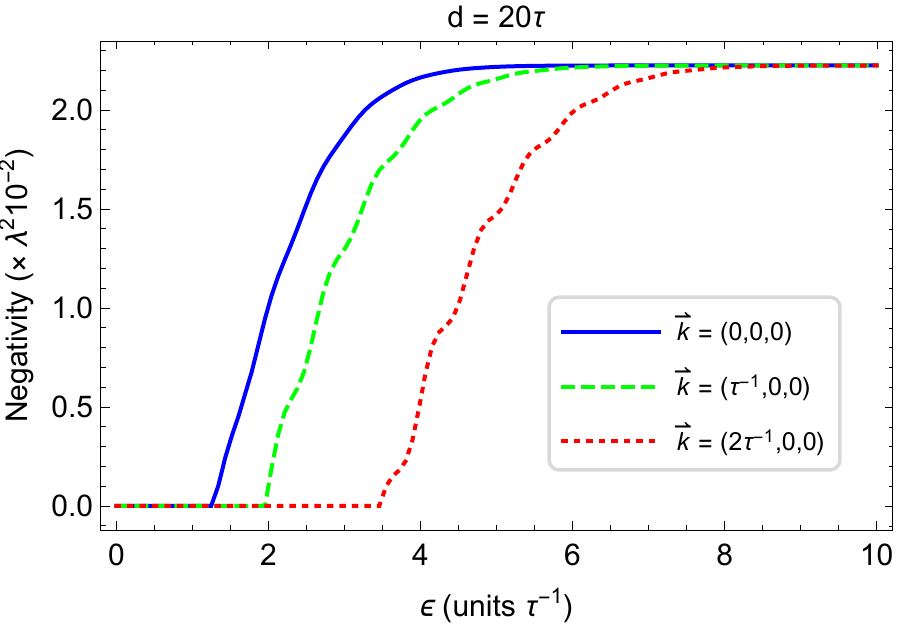}}
	\caption{Negativity of identical detectors as a function of the bandwidth of $\epsilon$ of modes squeezed, centered around a mode $\bm k$. The squeezing inside the bandlimited range is of uniform amplitude $r=1$, and outside is zero. The detectors are coupled to the field through Gaussian switching functions of width $\uptau$ centered at $t=0$, they are centered at $(\pm d/2,0,0)$ and have Gaussian spatial profiles of width $\sigma=\uptau$, and their energy gaps are $\Omega=\uptau^{-1}$.}
	\label{fig:neg_vs_epsilon}
\end{figure}

In Fig.~\ref{fig:neg_vs_epsilon} we plot the dependence of the negativity that two UDW detectors can harvest from the field, as a function of the bandwidth $\epsilon$ of field modes that are squeezed (we assume the squeezed modes to be centered around some wavevector $\bm k$). In the top plot of this figure, we suppose that the detectors are near enough in space such that they are able to harvest entanglement from the field vacuum ($\epsilon=0$). Perhaps unintuitively, we find that as we start squeezing around the mode $\bm k$ (i.e. we increase $\epsilon$), the negativity of the detectors initially begins to decrease. That is, for a small bandwidth $\epsilon$ of field squeezing, regardless of the mode $\bm k$ around which the squeezing is being performed, the amount of entanglement that the detectors can harvest from the field is actually less than what they could harvest from the vacuum. Eventually however, as the bandwidth is increased further, the amount of entanglement that the detectors can harvest from the field becomes higher than in the vacuum case. 

Meanwhile, detectors with a large spatial separation (bottom plot of Fig.~\ref{fig:neg_vs_epsilon}) are unable to harvest entanglement from the vacuum ($\epsilon=0$), as was already shown in Ref.~\cite{Pozas2015}. In this case increasing the squeezing bandwidth allows the detectors to harvest some entanglement, but this only occurs for $\epsilon$ larger than some critical value $\epsilon_c$. Hence, regardless of separation, the ability of a pair of UDW detectors to harvest more entanglement from a squeezed field state than from the vacuum is dependent on whether a large enough frequency interval of field modes is squeezed, i.e. if the bandwidth $\epsilon$ is larger than some critical value $\epsilon_c$.

We notice from the plots in Fig.~\ref{fig:neg_vs_epsilon} that the critical bandwidth $\epsilon_c$ necessary to achieve an improvement in entanglement harvesting over the vacuum is at least of the order $|\bm k|$, where $\bm k$ is the wavevector of the mode around which we squeeze. Hence for instance if we wanted to use a 300 THz squeezed laser source to entangle a pair of atomic detectors, we would need to squeeze all the modes up to 600 THz with wavevectors pointing in the direction of the laser, as well a wide range of field modes pointing in other directions. As far as we are aware, current experimental setups featuring squeezed electromagnetic field states do not squeeze such large bandwidths of field modes. Hence, in order to make use of the benefits of squeezed field states with respect to entanglement harvesting, it may be necessary to increase the experimentally achievable squeezing bandwidth. Alternatively, it might still be possible to obtain high levels of harvestable entanglement with narrowly bandlimited squeezed states, but for which the squeezing amplitude $\zeta(\bm k)$ is non-uniform in the bandlimited range. This remains to be investigated in future work.

\section{Conclusions}\label{sec:conclusions}

We studied the ability of a pair of Unruh DeWitt particle detectors to harvest quantum and classical correlations from thermal and squeezed states of a scalar field with which they interact. We find several interesting results:

First, we prove that the amount of entanglement that a pair of identical detectors (with arbitrary spatial profiles and time-dependent switching functions) can harvest from a thermal state of the field decreases monotonically with  temperature. Additionally, we obtain a lower bound on this rate of decrease, and hence show that for temperatures higher than a certain threshold the detectors are unable to harvest any entanglement from the field. With these findings we also extend the main results in \cite{Brown2013a}, where it was numerically shown (using the very different formalism of Gaussian quantum mechanics) that temperature is detrimental to entanglement harvesting by harmonic oscillator detectors from a massless field in 1+1 dimensional spacetime. Indeed, we prove that this is also the case for qubit detectors of arbitrary shape and switching interacting with a field of any mass in any dimensionality of spacetime.

On the other hand, we find that unlike the negativity, the mutual information --- which is a measure of the total (quantum and classical) correlations --- that the detectors harvest from the field actually increases linearly with the field temperature (again extending the numerical findings of \cite{Brown2013a} to qubit detectors). Hence, while thermal noise hinders the ability of UDW detectors to harvest entanglement, it is beneficial in the harvesting of non-entanglement correlations.

Moving on to squeezed field states, we start by proving that, at least to leading perturbative order, the amount of entanglement that a UDW detector pair can harvest from a squeezed coherent state is independent of its coherent amplitude. This greatly generalizes the result of Ref.~\cite{Simidzija2017b}, which considered only unsqueezed coherent states, to hold for all general squeezed coherent states.

We also show that, unlike the coherent amplitude, the field's squeezing amplitude $\zeta(\bm k)$ \textit{does} affect the amount of entanglement that the detectors can harvest from the field. In particular, we find that the amount of entanglement that detectors centered at a spatial point $\bm x_0$ can harvest is directly related to the amplitude of the Fourier transform of $\zeta(\bm k)$ evaluated at $\bm x_0$. Hence, contrary to vacuum~\cite{Pozas2015}, coherent~\cite{Simidzija2017b}, and thermal states,  harvesting entanglement from general squeezed states is generally \textit{not} a translationally invariant process.

However, and perhaps surprisingly, we find that for detectors centered at a particular location $\bm x_0$, the amount of entanglement harvested from a highly and uniformly squeezed state is independent of the spatial separation of the detectors. Moreover, this amount of entanglement is often much larger than detectors at the same separation would be able to harvest from the vacuum, raising the idea of the possibility of using squeezed states to experimentally test entanglement harvesting. This result is commensurate with the finding that squeezed states can remove the distance decay of protocols that rely on field entanglement such as quantum energy teleportation \cite{Hotta2014}. 

Finally, we have also studied how entanglement harvesting is modified when we allow for squeezing only in a finite frequency bandwidth of field modes. We find that if we restrict the modes of the field that are squeezed to a narrow bandwidth (namely, when the bandwidth is below the order of the frequency being squeezed), then squeezing states give no noticeable advantage over vacuum entanglement harvesting, at least for uniform squeezing. It remains to be seen whether a more general squeezing amplitude (e.g. with continuously varying magnitude and phase) can provide the necessary advantages in entanglement harvesting that we have found here for uniform squeezing, while at the same time being implementable in a lab setting. This is an important direction for future research, since such a squeezed field state could overcome the main experimental limitation of entanglement harvesting: the fast decay with detector separation.

\acknowledgements

P.S. gratefully acknowledges the support of the NSERC CGS-M and Ontario Graduate Scholarships. E.M.-M. acknowledges the funding from the NSERC Discovery program and his Ontario Early Research Award.

\appendix

\section{Thermal two-point function}\label{app:thermal_wightman}

We will show that our expression for the thermal two-point function in Eq.~\eqref{eq:thermal_wightman} reduces to the special case in Eq.~\eqref{eq:thermal_wightman_alt} when $m=0$, $n=3$, and $\bm x' = t' = 0$.

Let us first evaluate the second term in Eq.~\eqref{eq:thermal_wightman}, $w_\beta(\bm x,t,0,0)$, which is given in Eq.~\eqref{eq:wbeta}. Working in polar coordinates, with $k:=|\bm k|$ and $r:=|\bm x|$, we straightforwardly obtain
\begin{align}
\label{eq:wbeta2}
    w_\beta(\bm x,t,0,0)
    &=
    \frac{1}{2\pi^2 r}\int_0^\infty
    \frac{\dif k}{e^{\beta k}-1}
    \sin(kr)\cos(kt)\notag\\
    &=
    \mathcal P\Bigg(
    -\frac{1}{4\pi^2(r^2-t^2)}
    \\
    &
    +\!\frac{1}{8\pi r\beta}\!
    \!\left[\coth\!\!\left(\frac{\pi(r+t)}{\beta}\!\right)\!\!+\coth\!\!\left(\frac{\pi(r-t)}{\beta}\!\right)\!\!
    \right]\!\!
    \Bigg)\!,
    \notag
\end{align}
where $\mathcal P$ denotes the principal value of the integral (this expression only has meaning as a distribution). Interestingly, notice that the last term does not depend on the temperature.

We can similarly calculate the first term in Eq.~\eqref{eq:thermal_wightman}, $w_0(\bm x,t,0,0)$, which is given in Eq.~\eqref{eq:wvac}. We obtain
\begin{align}
    w_0(\bm x,t,0,0)
    &=
    \frac{1}{8\pi^2\ii r}
    \int_0^\infty\!\!\dif k
    \left(
    e^{-\ii k(t-r)}-e^{-\ii k(t+r)}
    \right)\notag\\
    &=
    \frac{1}{8\pi^2\ii r}
    \lim_{s\rightarrow\infty}
    \int_0^s\!\!\!\dif k
    \left(\!
    e^{-\ii k(t-r)}-e^{-\ii k(t+r)}
    \!\right)\notag\\
    &=
    \mathcal P\left(\frac{1}{4\pi^2(r^2-t^2)}\right)
    +\lim_{s\rightarrow\infty}
    \frac{1}{8\pi^2 r} \label{eq:w0_intermediate}\\
    &\phantom{=}\times
    \Bigg[\frac{\ii\sin\left(s(r+t)\right)}{r+t}-\frac{\ii\sin\left(s(r-t)\right)}{r-t}
    \notag\\
    &\phantom{===}-
    \frac{\cos\left(s(r+t)\right)}{r+t}-\frac{\cos\left(s(r-t)\right)}{r-t}
    \Bigg].\notag
\end{align}
Notice that although these limits do not converge as real functions, they do converge as distributions on test functions. Namely we have
\begin{align}
    \lim_{s\rightarrow\infty}
    \frac{\sin(s x)}{\pi x}&=
    \delta(x), \\
    \lim_{s\rightarrow\infty}
    \frac{\cos(s x)}{\pi x}&=
    0 =\text{the zero distribution}.
\end{align}
Hence Eq.~\eqref{eq:w0_intermediate} simplifies to
\begin{align}
\label{eq:w0_2}
    w_0(\bm x,t,0,0)
    =&
    \mathcal P\left(\frac{1}{4\pi^2(r^2-t^2)}\right)
    \notag\\
    &+
    \frac{\ii}{8\pi r}
    \left[\delta^{(3)}(r+t)-\delta^{(3)}(r-t)]
    \right],
\end{align}
where it should again be emphasized that the principal value and the delta functions only make sense as distributions. Finally, combining Eqs.~\eqref{eq:wbeta2} and \eqref{eq:w0_2}, we find that for a massless field in $(3+1)$-dimensions our expression for the two-point function, Eq.~\eqref{eq:thermal_wightman},  reduces to the distribution in Eq.~\eqref{eq:thermal_wightman_alt}, which was obtained in~\cite{Weldon2000} by a completely different method.

\twocolumngrid

\bibliography{references}
\bibliographystyle{apsrev4-1}
	
\end{document}